\documentclass{aa}
\usepackage{graphicx}
\usepackage{txfonts}
\usepackage{natbib}

\usepackage{url}
\newcommand{\hii}{{\sc Hii}}
\newcommand{\mic}{~$\mu$m}

\begin{document}
   \title{ATLASGAL - The APEX Telescope Large Area Survey of the Galaxy at
   870\mic}

   \author{F.~Schuller\inst{1} \and
     K.~M.~Menten\inst{1} \and Y.~Contreras\inst{1,2}
     \and F.~Wyrowski\inst{1}  \and P.~Schilke\inst{1}
     \and L.~Bronfman\inst{2} \and
     T.~Henning\inst{3} \and C. M.~Walmsley\inst{4} \and
     H.~Beuther\inst{3} \and S.~Bontemps\inst{5} \and
     R.~Cesaroni\inst{4} \and
     L.~Deharveng\inst{6} \and G.~Garay\inst{2} \and
     F.~Herpin\inst{5} \and B.~Lefloch\inst{7} \and
     H.~Linz\inst{3} \and
     D.~Mardones\inst{2} \and V.~Minier\inst{8} \and
     S.~Molinari\inst{9} \and F.~Motte\inst{8} \and
     L.-\AA.~Nyman\inst{10} \and V.~Reveret\inst{10}
     \and C.~Risacher\inst{10} \and D.~Russeil\inst{6}
     \and N.~Schneider\inst{8} \and L.~Testi\inst{11}
     \and T.~Troost\inst{1} \and T.~Vasyunina\inst{3}
     \and M. Wienen\inst{1} \and A.~Zavagno\inst{6}
     \and A.~Kovacs\inst{1} \and E.~Kreysa\inst{1}
     \and G.~Siringo\inst{1} \and A.~Wei\ss\inst{1}
   }

   \offprints{F.~Schuller}

   \institute{Max-Planck-Institut f\"ur Radioastronomie, Auf dem H\"ugel 69,
     D-53121 Bonn, Germany\\
     \email{schuller@mpifr-bonn.mpg.de}
     \and
     Departamento de Astronom\'ia, Universidad de Chile,
     Casilla 36-D, Santiago, Chile
     \and
     Max-Planck-Institut f\"ur Astronomie, K\"onigstuhl 17,
     D-69117 Heidelberg, Germany
     \and
     Osservatorio Astrofisico di Arcetri, Largo E. Fermi, 5,
     I-50125 Firenze, Italy
     \and
     Laboratoire d'Astrophyisique de Bordeaux - UMR 5804, CNRS - Universit\'e
     Bordeaux 1, BP 89, F-33270 Floirac, France
     \and
     Laboratoire d'Astrophysique de Marseille - UMR 6110, CNRS - Universit\'e
     de Provence, 
     F-13388, Marseille Cedex 13,
     France
     \and
     Laboratoire d'Astrophysique de l'Observatoire de Grenoble,
     BP 53, F-38041 Grenoble Cedex 9, France
     \and
     Laboratoire AIM, CEA/IRFU - CNRS - Universit\'e Paris Diderot, Service
     d'Astrophysique, F-91191 Gif-sur-Yvette, France
     \and
     Istituto Fisica Spazio Interplanetario - INAF,
     Via Fosso del Cavaliere 100, I-00133 Roma, Italy
     \and
     ESO, Alonso de Cordova 3107, Casilla 19001, Santiago 19, Chile
     \and
     ESO, Karl Schwarzschild-Strasse 2, D-85748 Garching bei M\"unchen, Germany
   }

   \date{Received 23 December 2008; accepted 2 March 2009}


  \abstract
  {Thanks to its excellent 5100~m high site in Chajnantor, the Atacama
  Pathfinder Experiment (APEX) systematically explores the southern sky at
  submillimeter wavelengths, in both continuum and spectral line emission.
  Studying continuum emission from interstellar dust is essential to locating
  the highest density regions in the interstellar medium, and deriving their
  masses, column densities, density structures, and large-scale morphologies.
  In particular, the early stages of (massive) star formation remain
  poorly understood, mainly because only small samples of high-mass
  proto-stellar or young stellar objects have been studied in detail so far.}
  {Our goal is to produce a large-scale, systematic database of massive
  pre- and proto-stellar clumps in the Galaxy,
  to understand how and under what conditions star
  formation takes place. Only a systematic survey
  of the Galactic Plane can provide the statistical basis for
  unbiased studies. A well characterized sample of Galactic star-forming
  sites will deliver an evolutionary sequence and a mass function of
  high-mass, star-forming clumps. This systematic survey at
  submillimeter wavelengths also represents a preparatory work
  for Herschel and ALMA.}
  {The APEX telescope is ideally located to observe the inner Milky Way.
  The Large APEX Bolometer Camera (LABOCA)
  is a 295-element bolometer array observing at 870\mic, with a beam
  size of $19\farcs2$. Taking advantage of its large
  field of view (11$\farcm4$) and excellent sensitivity, we started an
  unbiased survey of the entire Galactic Plane accessible to APEX, with
  a typical noise level of 50--70~mJy/beam: the APEX Telescope Large Area
  Survey of the Galaxy (ATLASGAL).}
  {As a first step, we covered $\sim$95~deg$^2$ of the Galactic Plane.
  These data reveal $\sim$6000 compact sources brighter than 0.25~Jy,
  or 63 sources per square degree, as well as extended
  structures, many of them filamentary. About two thirds of the compact
  sources have no bright infrared counterpart, and some of them are
  likely to correspond to the precursors of
  (high-mass) proto-stars or proto-clusters. Other compact sources
  harbor hot cores, compact \hii ~regions, or young embedded clusters,
  thus tracing more evolved stages after massive stars have formed.
  Assuming a typical distance of 5~kpc, most sources are clumps smaller
  than 1~pc with masses from a few 10 to a few 100~M$_{\sun}$. In
  this first introductory paper, we show preliminary results from these
  ongoing observations, and discuss the mid- and long-term
  perspectives of the survey.}
  {}


   \maketitle
%

\section{Introduction}

Dust continuum emission in the (sub)millimeter (submm) range is one of
the most reliable tracers of the earliest phases of star formation since
it directly probes the dense interstellar material from which the stars
form. Trying to understand the formation and early evolution of stars is
an important field of modern astrophysics \cite[see the review of][and
references therein]{McKee2007}. 

A large amount of theoretical effort has been
aimed at constraining the early stages of the formation of isolated
low mass (solar-like) stars, although two quite different
pictures are being maintained. On the one hand, a collapse picture has
emerged whose ``long'' timescales are determined by ambipolar diffusion
\citep{shu+1987,adams+1987}.
On the other, a scenario is advocated in which the star formation rate is
governed by supersonic turbulence \citep{maclowklessen2004}.
Many observational campaigns have contributed crucial data and
in particular systematic (sub)millimeter surveys of nearby star-forming
regions, such as Chamaeleon, Perseus, and Ophiuchus, have resulted in the
detection of many prestellar and proto-stellar condensations
\citep{andre-pp4,ward-pp5}. They have allowed determination of the
dense core mass spectrum down to substellar masses, and investigation
of its relationship to the initial mass function \citep{motte1998,
gahm2002, hatchell2005}.

High-mass star formation is even less well constrained
\citep[e.g.,][for reviews]{ZinneckerYorke,beuther+2007} and there
is controversy about how these stars attain their final masses:
either driven by turbulence \citep{mckee-tan-2003} or
competitive accretion \cite[e.g.,][]{bonnell2004}.
Since high-mass stars seem to form mostly in clusters (along with
many more lower-mass stars),
it is fundamental to have a good estimate of the mass distribution of
cluster-forming clumps in the Giant Molecular Clouds (GMCs) they are part
of. This is exactly one of the major goals of the present effort.

A cardinal reason for the many open questions concerning high-mass star
formation is that the relevant timescales are short \citep{mckee-tan},
implying low number statistics in finding stars in a given evolutionary
state. Moreover, high-mass stars are rare. This implies that regions in
which they form are generally at large distances, which leads to the basic
problem that there objects are difficult to identify. For instance, little
is known about the Galactic molecular ring, a few kpc distant from us,
where most of the star formation is presently taking place
\citep[e.g.,][]{bronfman2000}.

While various samples of high-mass proto-stellar objects have been
defined based on various criteria, such as maser emission
\citep[e.g.,][]{walsh+1997,pesta2005} and infrared (IR) colors
\citep{palla+1991, molinari+1996, sridharan+2002, lumsden+2002,
robitaille2008}, an \textit{unbiased} sample has not yet been assembled.
Much effort has focused on follow-up studies
of bright IR sources detected by the Infrared Astronomy Satellite
(IRAS) and showing far-infrared (FIR) colors typical of ultra-compact
\hii ~regions \citep{wood+churchwell1989}, e.g., in high-density
molecular tracers \citep{Bronfman1996,molinari+1996},
dust continuum emission \citep{faundez+2004},
or maser emission \citep[e.g.,][]{palla+1993,beuther-masers}.
While some studies have revealed a few massive cold cores
in the neighborhood of the central compact \hii ~region
\citep[e.g.,][]{garay+2004},
they were mostly biased against the earliest, possibly
coldest phases of massive star formation.

Alternatively, dense condensations in Infrared Dark Clouds
\citep[IRDCs, see][for a compilation]{simon+2006a}
are promising
hunting grounds for the initial stages of high-mass star formation
(e.g., \citealt{rathborne+2006}; see also the reviews by
\citealt{menten+2005,wyrowski2007}). IRDCs
appear in absorption by dust even at mid-infrared wavelengths against
the diffuse emission of the Galactic Plane and trace high column
densities. However, IRDCs become increasingly difficult to identify
the further away they are, and the relative contributions by foreground
and background emission may be difficult to estimate. Moreover, not
all massive dust condensations appear as infrared dark clouds.
For example, the two most massive condensations (with masses of 200 and
800~M$_{\sun}$ at a distance of 1.3~kpc) observed at 870\mic ~in the
photo-dissociation region (PDR) bordering the RCW~120 \hii ~region
\citep{ref-rcw120} are not detected as IRDCs at 8 or 24\mic.
They are hidden, at these wavelengths, by the bright emission of the
adjacent foreground PDR.

Cold dust absorbing the IR radiation also shows thermal emission in the submm
regime. This grey body emission is generally optically thin and, thus,
an excellent tracer of the dust mass and, by inference, total mass
of the emitting cloud. Therefore, observations in the submm continuum
are complementary to the IR extinction techniques.
Given the short timescales for the formation
of high-mass stars, all early formation stages will still be associated
with strong dust continuum emission, allowing a systematic study of a
large range of evolutionary phases.

Several groups have performed large-scale mapping of
molecular complexes in the (sub)millimeter continuum.
Using the  Submillimetre Common User Bolometer Array (SCUBA)
instrument, \citet{ref-johnstone} mapped 4~deg$^2$
of the Ophiuchus star-forming cloud, and \citet{hatchell2005}
covered 3~deg$^2$ of the Perseus molecular cloud. Complementary
to the Spitzer ``Cores to Disks'' legacy project \citep{ref-c2d},
\citet{enoch+2006} mapped 7.5~deg$^2$ covering the Perseus cloud,
and \citet{young+2006} mapped nearly 11~deg$^2$ covering Ophiuchus,
both using the BOLOCAM instrument at 1.1~mm.

In the field of high-mass star formation, \citet{ref-moore}
covered 0.9~deg$^2$ of the W3 giant molecular cloud with SCUBA.
Going to larger scales, \citet{motte+2007} used the MAMBO instrument
to obtain a complete dust-continuum map of the molecular complex
associated with the Cygnus X region. From this 3~deg$^2$ map, they
could derive some statistically significant results about
the characteristics (e.g., mass, density, outflow power, mid-IR flux)
and lifetimes of different evolutionary phases of high-mass star
formation. In particular, no high-mass analog of pre-stellar dense
cores was found by this study, hinting at a short lifetime
for the earliest phases of dense cores in which high-mass stars are being
formed. An unbiased survey of the complete Galactic Plane will allow us to
place far stronger constraints on this stage of high-mass star formation.

On an excellent 5100~m high site and at a latitude of $-23\degr$,
the 12~m diameter APEX telescope \citep{guesten+2006a,guesten+2006b}
is ideally located to make sensitive
observations of the two inner quadrants of the Galactic Plane.
A consortium led by the Max Planck Institute f\"ur Radioastronomie
(MPIfR) in Bonn, involving scientists from the Max Planck Institute f\"ur
Astronomie in Heidelberg and  from the ESO and Chilean communities
has embarked on the APEX Telescope Large Area Survey of the
Galaxy (ATLASGAL). This project aims at a systematic survey of the inner
Galactic Plane, mapping several 100~deg$^2$ with a uniform
sensitivity. In this paper, we present the first data taken and analyzed.
The observing strategy and data reduction are described in
Sect.~\ref{sec-obs} and Sect.~\ref{sec-reduc}. First preliminary
results are discussed in Sect. \ref{sec-result}. Finally, we
discuss longer term perspectives in Sect. \ref{sec-persp}.


\section{\label{sec-obs}Observations}

\begin{figure*}
\centering
\includegraphics[width=\linewidth]{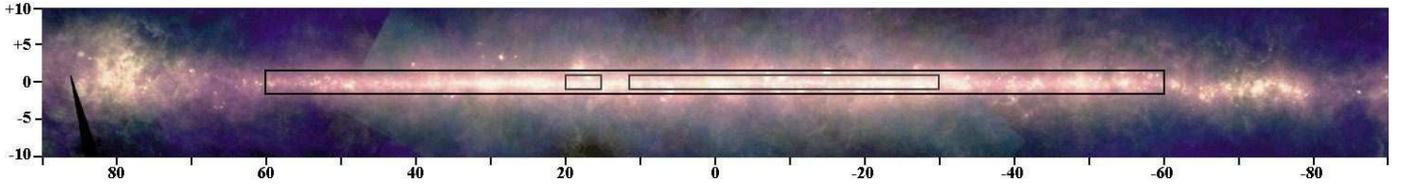}
\caption{Coverage of the ATLASGAL observations overlaid on an
  IRAS false color image (12\mic ~in blue, 60\mic ~in green
  and 100\mic ~in red). This image covers $\pm 90\degr$
  in galactic longitude, and $\pm 10\degr$ in latitude.
  The long frame shows the area that we plan to cover until
  the end of 2009: $\pm 60\degr$ in $l$ and $\pm 1.5\degr$ in $b$.
  The smaller frames inside delineate the area that was
  observed in 2007 (see also Fig.~\ref{fig-all-data}).}
\label{fig-coverage}
\end{figure*}

Observations of continuum emission from interstellar dust at
(sub)millimeter wavelengths are completed most effectively with sensitive
broadband bolometer detectors. Modern (sub)millimeter bolometer development
started in the mid 1980s with single-pixel-element instruments,
with a resolution, i.e., diffraction-limited FWHM beam size, of $11''$
for the IRAM 30~m telescope at 1.2~mm and $19''$ for the James Clerk
Maxwell Telescope (JCMT) at 1.1~mm. Since then,
arrays with more and more elements have been developed and deployed
to telescopes. Modern arrays of bolometers have hundreds of
elements: 384 for the Nyquist-sampled Submillimeter High-Angular
Resolution Camera~2 \citep[SHARC~II,][]{ref-sharc2}, and 295 for the
Large APEX Bolometer Camera (LABOCA), which was recently deployed
at the APEX telescope \citep{siringo2007,siringo2008}.
With a field of view of $11\farcm4$, and a single-pixel sensitivity
(noise equivalent flux density, NEFD) in the range 40--70~mJy~s$^{1/2}$,
LABOCA is the most powerful bolometer
array for large-scale mapping operational on a ground-based telescope.

The ATLASGAL observations are carried out at the APEX 12~m submm antenna,
which is a modified copy of the VERTEX prototype antenna for the Atacama
Large Millimeter Array (ALMA). The antenna has a surface accuracy
of 15\mic ~rms, i.e., higher than ALMA specifications.
Two additional Nasmyth cabins allow several receivers to be ready for use.
The LABOCA instrument is an array of 295 bolometers arranged
in an hexagonal pattern, with two-beam spacing between bolometers
\citep{siringo2008}. It is located at the Cassegrain focus,
where it has an effective field of view of $11\farcm4$ in diameter.
Small variations in the beam shape are seen within the field
of view, but should not affect the data \citep[see][for a
discussion of these effets]{siringo2008}. Its bandpass is centered
on 870\mic ~(frequency 345~GHz) with a bandwidth of 60~GHz.
The beam at this wavelength was measured to have a width of
$19\farcs2$ FWHM. The number of usable bolometers at the end of the
reduction, after flagging the noisy ones and the ones that do not
respond (Sect.~\ref{sec-reduc}) is usually between 250 and 260.

\begin{figure}
\centering
\resizebox{8.5cm}{!}{
\includegraphics*[20pt,10pt][770pt,400pt]{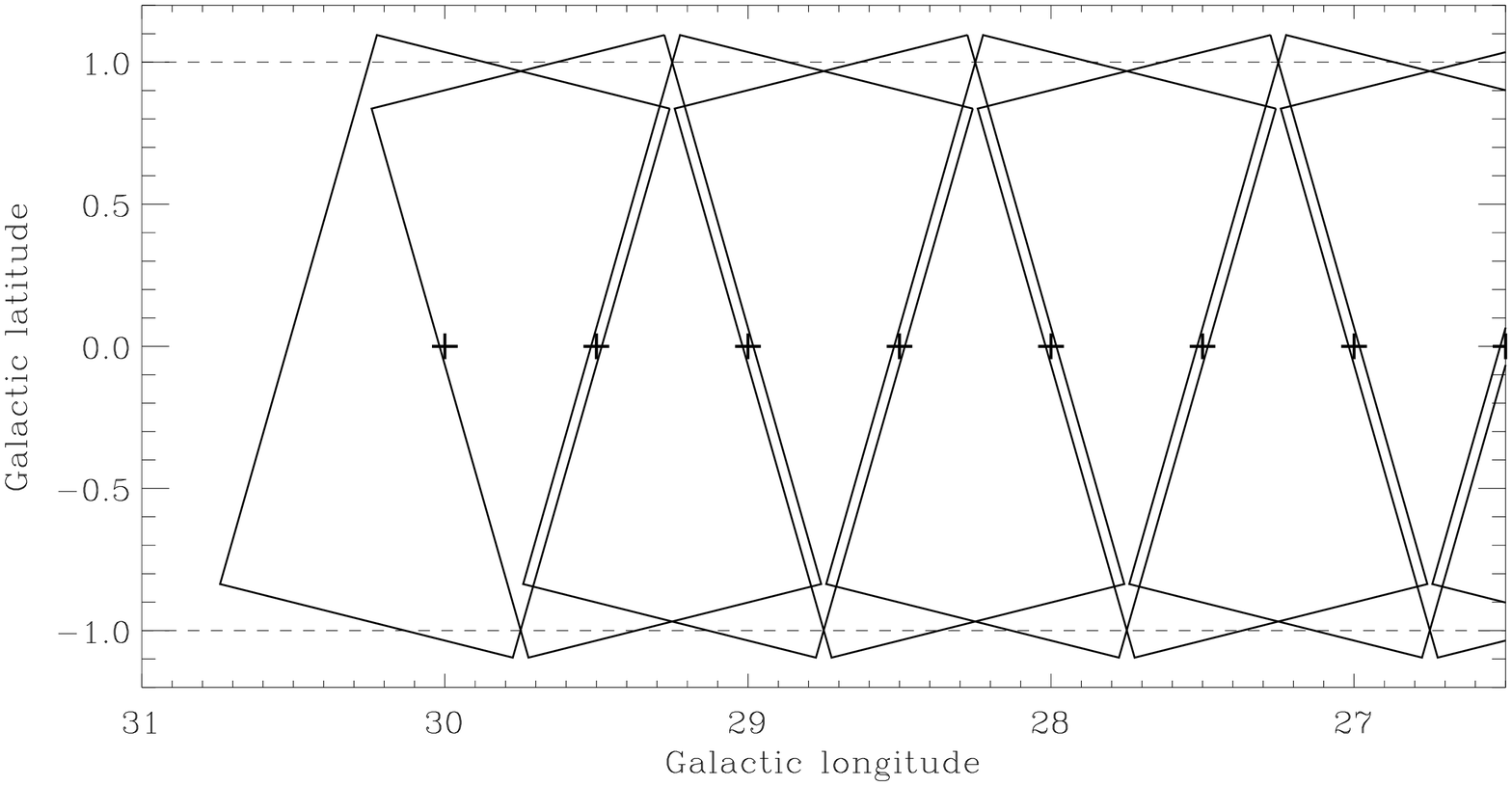}}
\caption{Mapping strategy: each rectangular frame represents a single
map. The plus symbols show the maps centers. A position angle of $15\degr$
with respect to the Galactic axis was chosen. Two neighboring maps are
observed with position angles of $+15\degr$ and $-15\degr$, respectively.}
\label{fig-observing}
\end{figure}

In the present paper, we focus on the data acquired in 2007, which cover
95~deg$^2$: $-30\degr \leq l \leq +11.5\degr$ and $+15\degr \leq l \leq
+21\degr$, with $\vert b \vert \leq 1\degr$ (Fig.~\ref{fig-coverage}),
at a one-$\sigma$ sensitivity of 50 to 70~mJy/beam.
For comparison, the 9-year lifetime of the SCUBA
instrument resulted in a total of 29.3~deg$^2$
mapped at 850\mic ~\citep{scuba-legacy}.
A single observation consists in an on-the-fly map, $2\degr$ long and
$1\degr$ wide, with $1\farcm5$ steps between individual lines, resulting
in fully sampled maps covering 2~deg$^2$. The scanning speed used is
3$'$/s. This fast scanning allows us to perform observations in total
power mode, i.e., without chopping. Each map
crosses the Galactic Plane with a $\pm15\degr$ position angle with
respect to the Galactic latitude axis. Two consecutive maps are
spaced by $0.5\degr$ along the Galactic Plane, with a $30\degr$
difference in position angle (Fig.~\ref{fig-observing}).
Thus, each position on the sky is observed in two independent maps,
of different position angles, greatly reducing many systematic
effects (e.g., striping in the scanning direction).

In the longer term, we plan to extend the coverage to $\pm60\degr$ in
longitude and $\pm1.5\degr$ in latitude (Fig.~\ref{fig-coverage}), with
the same sensitivity of 50~mJy/beam. An ESO large programme was approved
to conduct these observations, which will require a total of 430 hours,
including all overheads, to cover 360~deg$^2$. The observing time will be
shared between 45 hours from Chile, 225 hours from the Max-Planck-Society
(MPG), and 160 hours from ESO, spanning four semesters in 2008--2009.


\section{\label{sec-reduc}Data reduction and source extraction}

The raw data are recorded in MB-FITS (Multi-Beam FITS)
format by the FITS writer, which
is part of the APEX Control System \citep[APECS,][]{muders+2006}.
These data were processed using the
BOlometer array data Analysis package (BOA; Schuller et al. in prep.),
which was specifically developed as part
of the LABOCA project to reduce data obtained with bolometers
at the APEX telescope.
The reduction steps involved in the processing of the ATLASGAL data
are described in the next subsections.

\subsection{\label{sec-reduc1}First step: from raw data to 2~deg$^2$ maps}

Each single observation is processed and calibrated to compile a map
covering 2~deg$^2$. The data are calibrated by applying an opacity
correction, as determined from skydips observed typically every two
hours \citep[see][for more details]{siringo2008}. In addition, the flux
calibration is regularly (every hour) checked against primary calibrators
(the planets Mars, Neptune and Uranus) or secondary calibrators (bright
Galactic sources, for which the fluxes at 870\mic ~were measured during
the commissioning of LABOCA). If any discrepancy is found between the
opacity-corrected flux of a calibrator and its expected flux, then the
correction factor required to obtain the expected flux is also applied
to the science data. Given the uncertainties in the fluxes of
the planets themselves, we estimate that, altogether, the typical
calibration error should be lower than 15\%. Pointing scans are
also completed using bright sources every hour; the pointing rms
is typically of order 4$''$.

The reduction steps are as follows:
\begin{itemize}
\item
flagging bad pixels (those that do not respond, or do not see the sky)
\item
correlated noise removal (see below for details)
\item
flagging noisy pixels
\item
despiking
\item
low-frequency filtering (see below)
\item
first-order baseline on the entire timestream
\item
building a map, using natural weighting: each data point has a weight
1/rms$^2$, where rms is the standard deviation of each pixel
\end{itemize}

\begin{figure*}
\centering
\resizebox{\linewidth}{!}{\includegraphics{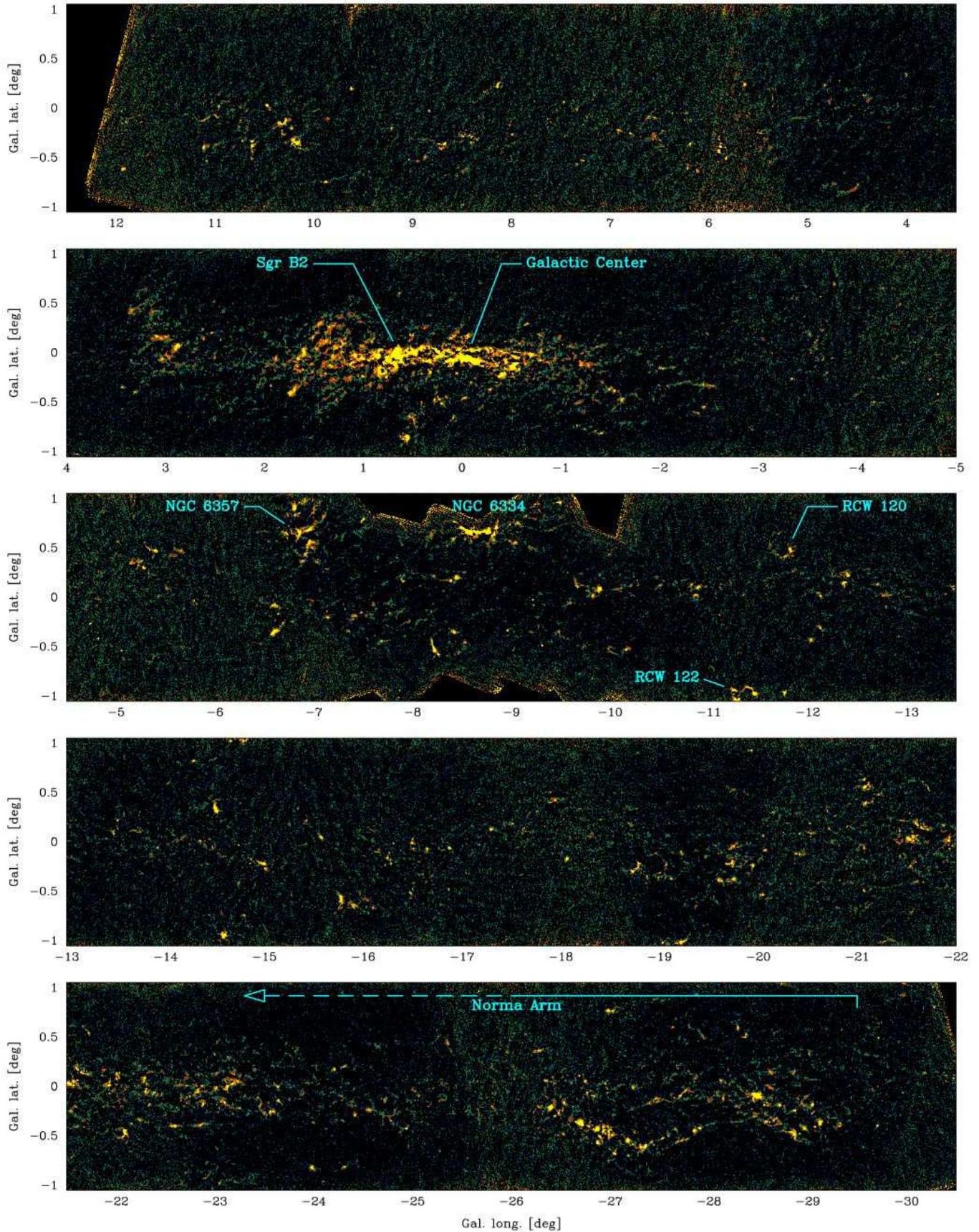}}
\caption{Combined maps showing 84~deg$^2$ of the ATLASGAL data
obtained in 2007, at 870\mic ~with a resolution of 19$''$.
The color scale is logarithmic between 0.05 and 1.0~Jy/beam. The rms is of
order 50 to 70~mJy/beam in most parts of these maps. A few known regions
are labeled in the maps. See also Fig.~\ref{fig-zooms}.}
\label{fig-all-data}
\end{figure*}

\begin{figure*}
\centering
\rotatebox{270}{
\includegraphics[viewport=22 80 555 690,width=58mm,clip]{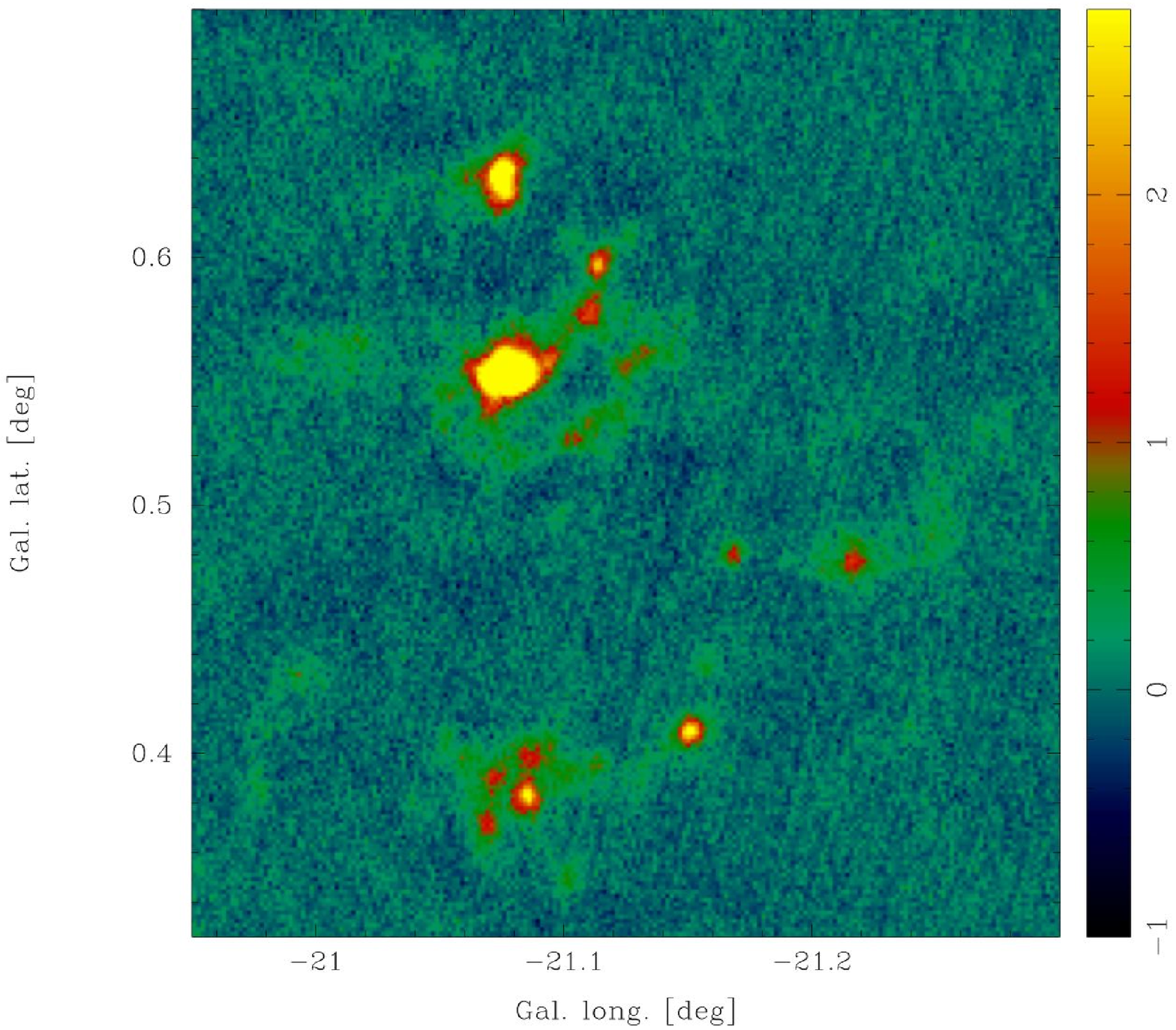}}
\rotatebox{270}{
\includegraphics[viewport=22 190 555 625,width=58mm,clip]{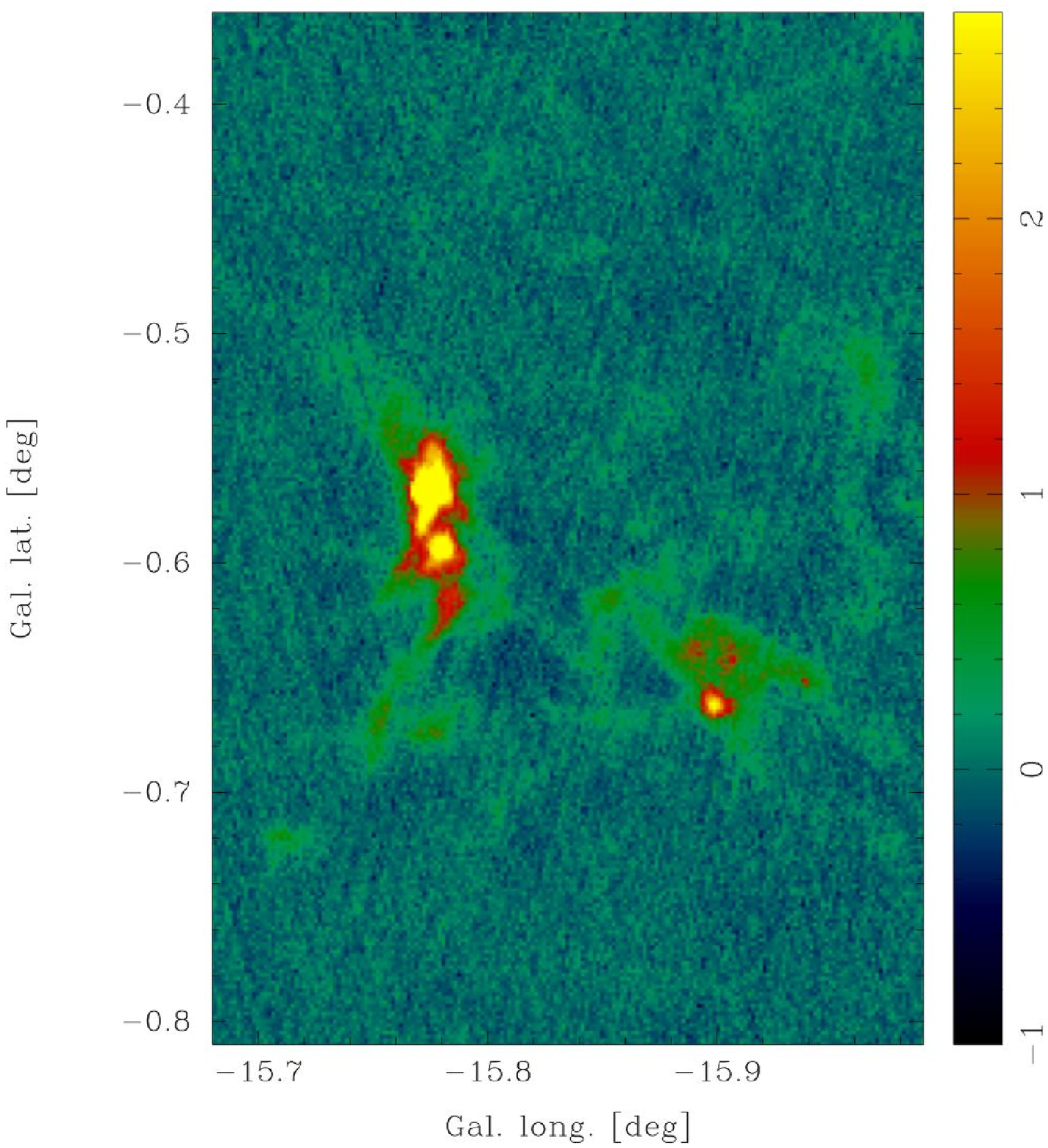}}
\rotatebox{270}{
\includegraphics[viewport=22 100 555 715,width=58mm,clip]{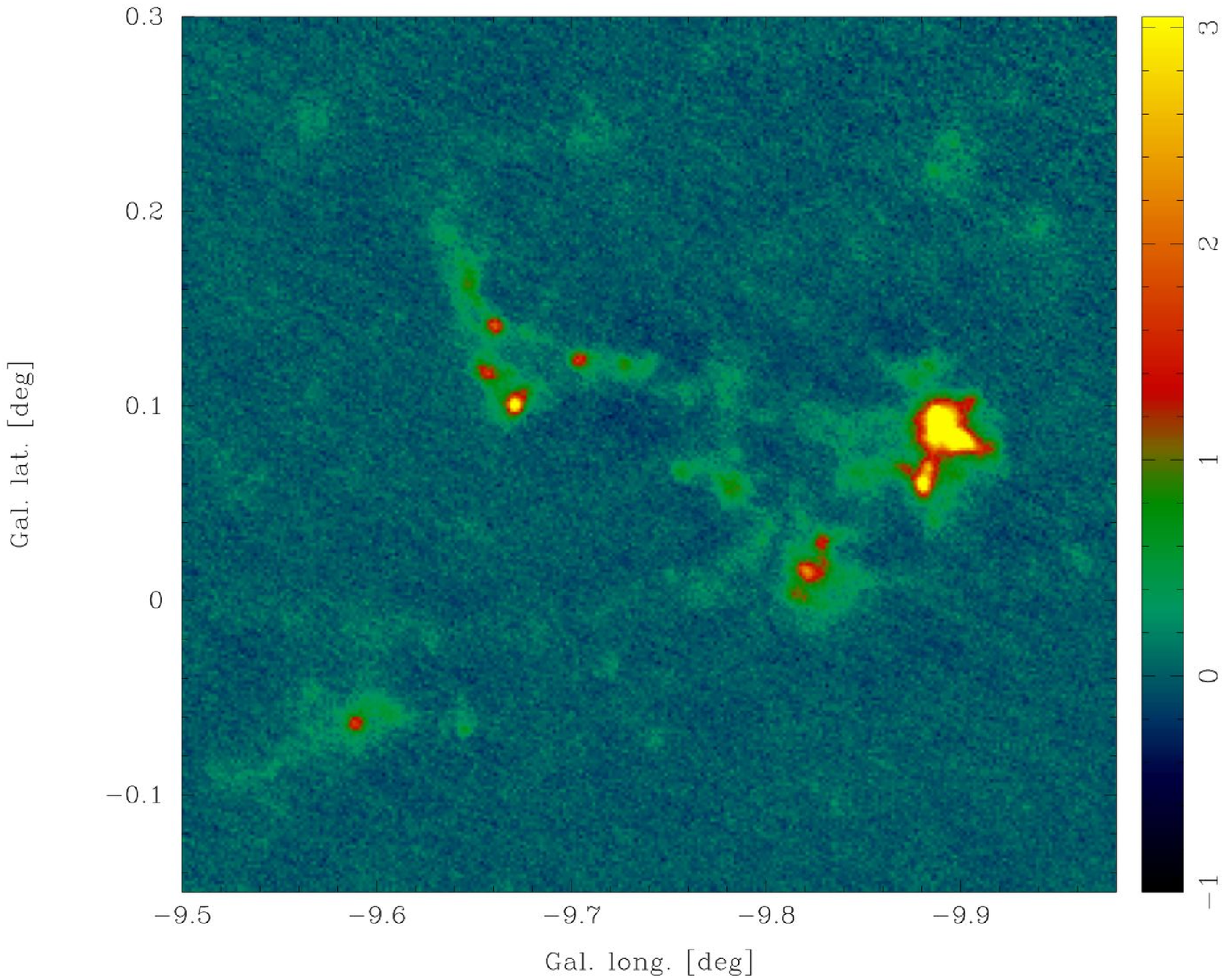}}
\caption{Zooms into selected regions. These maps are projected with three
pixels per beam, resulting in a pixel r.m.s.~of between 80 and 130~mJy/beam.
All maps are shown in linear scale, calibrated to Jy/beam. The color
scale has been chosen to highlight the negative artifacts, but doesn't
cover the full range of fluxes in these images: the brightest
peaks are, from left to right, 18.6, 17.2, and 7.2~Jy/beam;
the most negative pixels are at -0.49, -0.54, and -0.43~Jy/beam,
respectively.}
\label{fig-zooms}
\end{figure*}

The main source of noise in the submm is variable atmospheric emission
(sky noise), which varies slowly with typical amplitudes of several 100~Jy.
Fortunately, this sky noise is highly correlated between bolometers.
For each integration, the correlated noise is computed as the median
value of all normalized signals. It is removed in successive steps:
as a first step, the median for all bolometers is computed,
and then the median for all pixels
connected to the same amplifier box (groups of 60 to 80 pixels);
finally, the correlated signal is computed for groups of 20 to 25
pixels sharing the same read-out cable. A consequence of this processing
is that uniformly extended emission is filtered out, since there is no way
of distinguishing extended astronomical signal observable by all bolometers
from correlated (e.g., atmospheric) noise. The strongest effect occurs
when computing the correlated noise in groups of pixels: these
groups typically cover $\sim1\farcm5 \times 5'$ on the array, and since
the correlated noise is computed as the median value of the signals,
any emission that is seen by more than half the bolometers in a given
group is subtracted. Thus, {\em uniform} emission on scales larger
than $\sim2\farcm5$ is filtered out.

Another low-frequency filtering is performed, to reduce the
effects of the slow instrumental drifts, which are uncorrelated
between bolometers. This filtering is performed in the Fourier
domain, and applies to frequencies below 0.2~Hz, which,
given the scanning speed of 3$'$/s, translates to spatial scales
of above 15$'$, i.e., larger
than the LABOCA field of view. Therefore, this step should not
affect extended emission that has not already been filtered-out
by the correlated noise removal.

This reduction is optimized for recovering compact sources with
high signal-to-noise ratio. Alternatively, the data can be reduced
without subtracting all components of correlated noise, which
allows one to recover more extended emission, but results in an
increased noise level in the final maps.

\subsection{Second step: using a source model}

The data processed in the first step are then combined into groups of
three maps: for each position $l_0$, the maps centered on $l_0$,
$l_0 + 0.5\degr$, and $l_0 - 0.5\degr$ are computed and coadded to obtain
a two-fold coverage of the central 1$\times$2~deg$^2$ area (see
Fig.~\ref{fig-observing}). This combined map is then used as a
source model for masking in the timestreams the data that correspond
to positions in the map where the emission is above some given
threshold. The full reduction is then performed again, involving the
same processing steps as described in the previous section with
less conservative parameters for despiking and correlated-noise
removal. During this step, datapoints corresponding to
strong emission in the model map are masked for the various
computations (baseline removal, despiking, subtraction of
correlated noise), but the results of these computations are
also applied to the masked data.

Finally, another iteration is performed, this time {\em subtracting}
the source model (resulting from the previous step) from the raw
data. In this case, all datapoints are used in the computations,
and processing steps that may affect strong sources (e.g., despiking)
can be performed almost without caution, since most of the true
emission has been subtracted from the signals. Before adding the
source model back into the reduced data, weights are computed using
sliding windows along the timestreams. The weight applied to each
datapoint is 1/rms$^2$, where rms is the standard deviation in a
given pixel computed for 50 datapoints.

The net effect of these
iterations with a source model is to reduce
the negative artifacts appearing around strong sources in the
first reduction step and to recover some fraction of the flux
of bright objects lost in that step.
The final flux uncertainty for compact sources should be lower
than 15\% (Sect.~\ref{sec-reduc1}), as also confirmed by
hundreds of measurements of primary
and secondary calibrators \citep{siringo2008}.

When compiling maps with two pixels per beam, the final noise
level obtained from individual observations is generally in
the 70--100~mJy/beam range. In the combined maps, the rms is then
50--70~mJy/beam in overlapping regions.
The combined maps covering continuously $-30\degr$ to $+12\degr$
in longitude are shown in Fig.~\ref{fig-all-data}, and
example zooms are shown in Fig.~\ref{fig-zooms}.

\subsection{\label{sec-extraction}Compact source extraction}

A preliminary extraction of compact sources was completed using the
SExtractor programme \citep{ref-sextractor}. The extraction was
performed in two steps: first, the peaks were detected on signal-to-noise
ratio maps; SExtractor was then used to compute the source peak
positions and fluxes, as well as their sizes. Since we were only
interested in compact sources when compiling this first catalog, one
constraint used was that the ratio of major to minor axes should be lower
than 4. Details of the method and the complete catalog itself will
be published in a forthcoming paper (Contreras et al. in prep).

\begin{figure}
\centering
\resizebox{8.5cm}{!}{\includegraphics{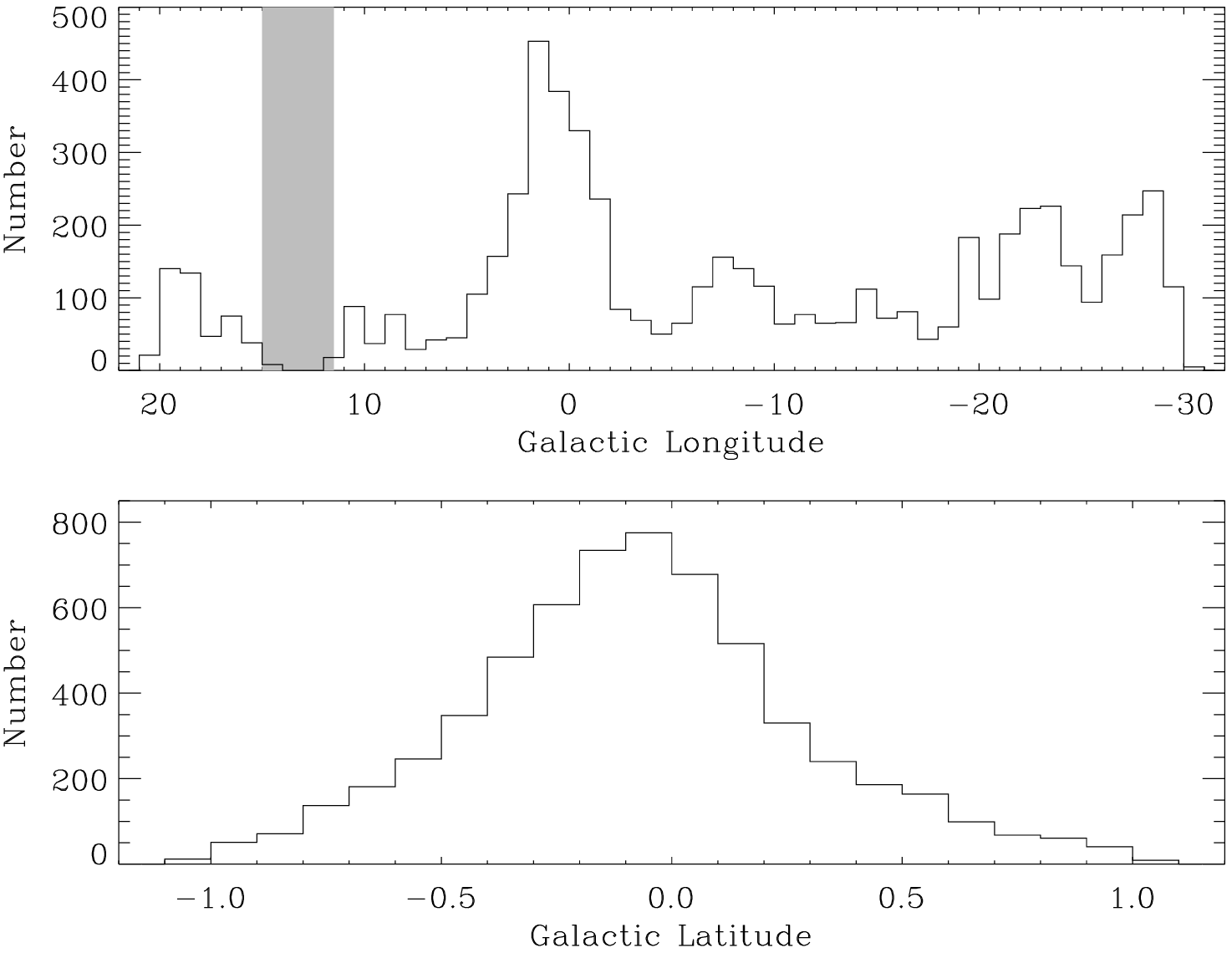}}
\caption{Distribution of the 6000 compact sources in Galactic
longitude (top, with 1$\degr$ bins) and latitude (bottom,
with 0.1$\degr$ bins). Note that the longitude range $+11.5\degr$
to $+15\degr$ is not covered by the present data.}
\label{fig-dist-gal}
\end{figure}

This source extraction resulted in $\sim$6000 objects with peak
fluxes above 250~mJy/beam, over the 95~deg$^2$ mapped.
The average source density is thus 63 sources per square degree, but
the source distribution is highly non-uniform (Fig.~\ref{fig-dist-gal}).
The source sizes range from one LABOCA beam size (19$''$) to about
8 beam widths (150$''$).
We note that the filtering of uniform emission on spatial scales
larger than $2\farcm5$ (Sect.~\ref{sec-reduc1}) should not prevent
extraction of sources of larger FWHM, as long as they are reasonably
compact, i.e., with a profile close to a Gaussian shape; their fluxes
however would be severely underestimated.

Assuming a typical distance of 5~kpc, which roughly corresponds to
the molecular ring, the 19$''$ resolution translates to $\sim$0.5~pc.
The LABOCA data are therefore sensitive to molecular clumps, following
the terminology of \citet[][see also \citealt{motte2008}]{williams+2000},
where dense cores have typical sizes of $\sim$0.1~pc and are embedded
in pc-scale clumps, themselves being part of $>$10~pc GMCs.
The sources distribution is strongly peaked toward the Galactic
Plane, as seen in Fig.~\ref{fig-dist-gal}, and in particular
in the direction of the Galactic Center, with an excess of sources
toward negative latitudes. Within the central few degrees, sources
exhibit an excess at positive longitudes.
Other peaks in longitude are visible in Fig.~\ref{fig-dist-gal}
in the direction of the Norma arm
($-30\degr \leq l \leq -20\degr$), and around $l = -8\degr$,
which corresponds to the NGC6334/NGC6357 star-forming complexes.

As seen in Fig.~\ref{fig-dist-gal}, the latitude distribution of
compact sources peaks at $b = -0.09\degr$, which is significantly
below zero. This effect is seen in all ranges of longitudes. We
suspect that this is due to the Sun being slightly above the Galactic
Plane, but this effect requires more study.
The reason for the asymmetry in
longitude within the few central degrees is unclear, but is certainly
real, as can be seen also in Fig.~\ref{fig-all-data}. Interestingly,
the distribution of compact sources seen with Spitzer at 24~$\mu$m
shows an excess toward {\em negative} longitudes \citep{ref-hinz}.
While no clear explanation is found, \citet{ref-hinz} suggested that
this could reflect the true distribution of infrared sources near the
Galactic Center, if more molecular clouds are present at positive
longitudes, obscuring the compact IR sources, which would be consistent
with the distribution that we observe in the submm.

\section{\label{sec-result}First results}

\begin{figure}
\centering
\includegraphics[width=\columnwidth]{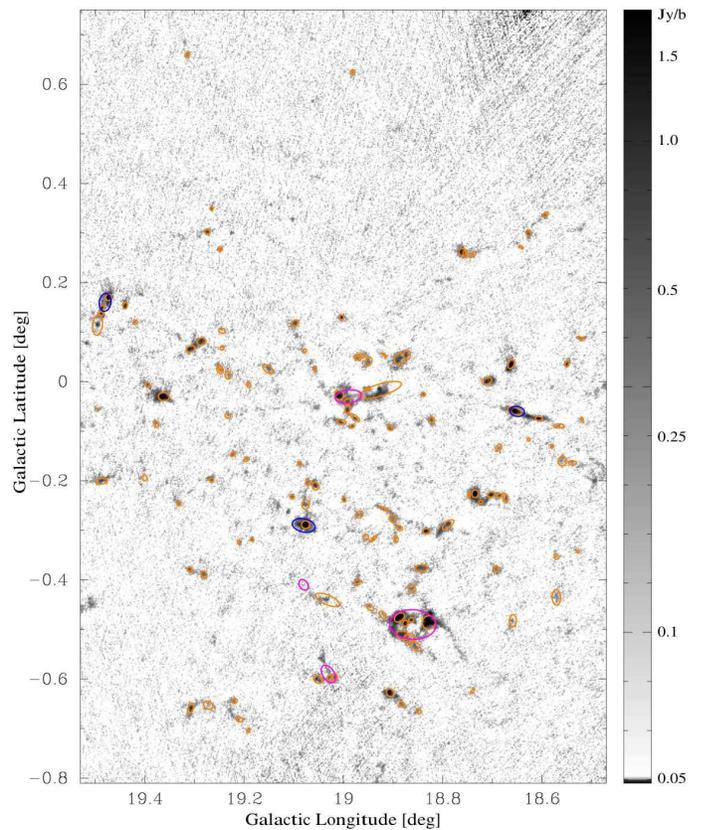}
\caption{Example combined map covering 1$\times$1.5~deg$^2$,
shown in logarithmic scale.
The pixel scale is one third of the beam and the
standard deviation between pixels in empty regions of the maps
is typically 90~mJy/beam. The brightest peak is at 6.2~Jy/beam.
The 128 compact sources extracted in this region are shown with
orange ellipses, with sizes and position angles as measured
with SExtractor (Sect.~\ref{sec-extraction}). Ultra-compact \hii
~regions discussed at the end of Sect.~\ref{sec-distance} are
shown with blue ellipses; compact and classical \hii ~regions are
indicated with magenta ellipses.}
\label{fig-exmap}
\end{figure}

The large scale maps in Fig.~\ref{fig-all-data} clearly show a variety
of  submillimeter sources from bright, compact objects to faint, extended
regions on the arcmin scale, and long filaments on almost a degree scale.
The survey is very rich, and a wealth of new results is expected. To
provide a flavor of these and to illustrate
how a complete view of the Galactic star-formation activity will
be achieved from ATLASGAL, we present a first analysis
and some first results for a typical $1\degr$ wide slice,
around $l = 19\degr$ (Fig.~\ref{fig-exmap}).
This region shows a reasonably rich population of sources,
and is also representative of regions with several cloud complexes 
at different distances. We also selected this region because it
is covered by the Galactic Ring Survey \citep[GRS,][]{jackson+2006},
which provides good quality CO data that we use to discuss the
association with large-scale molecular clouds.

\subsection{Compact sources and extended emission at $l = 19\degr$}

In a number of Galactic directions, there is submillimeter emission
virtually everywhere in the Galactic Plane. The region shown in
Fig.~\ref{fig-exmap} at $l = 19\degr$ is one of these. From our
preliminary source extraction, we identified 128 compact sources
with peak fluxes above 0.25~Jy/beam in this 1$\times$2~deg$^2$ slice
(no source was detected outside the $\vert b \vert \, \leq \, 0.75\degr$
range shown in Fig.~\ref{fig-exmap}),
in perfect agreement with the average 63 extracted sources per square
degree. Only 16 of these sources have a probable known counterpart
in the SIMBAD database (infrared, millimeter or radio source within
a 10$''$ search radius), which means that more than 80\% have not yet
been studied. Some of them certainly have counterparts in the new
infrared surveys observed with Spitzer (see e.g., Fig.~\ref{fig-color});
associations between ATLASGAL and Spitzer data will be addressed
elsewhere.

As can be seen in Fig.~\ref{fig-exmap}, most of the compact sources
are embedded in more extended emission of a few arcmin
extent. Many of these features appear filamentary and could
correspond to overdensities within giant molecular clouds, in
which even denser clumps are forming. That small groups of clumps
belong to one single molecular complex is sometimes confirmed by
information about their kinematics, as discussed in the next section.

\subsection{\label{sec-distance}Coherent complexes and distance determination}

Toward the inner Galactic Plane, the detected sources can be at any
distance. {\em A priori}, any direction of the Plane could contain sources
at distances distributed over a significant range of values, except for
some peculiar lines of sight, like the Galactic Center (GC), where most
sources are probably concentrated close to the GC itself. In other
directions, one expects to find overdensities of objects at particular
distances, corresponding to the Galactic arms, particularly in the
molecular ring region, but a wide range of distances is allowed.

\begin{figure}[bp]
\centering
\rotatebox{270}{\includegraphics[width=10.0cm]{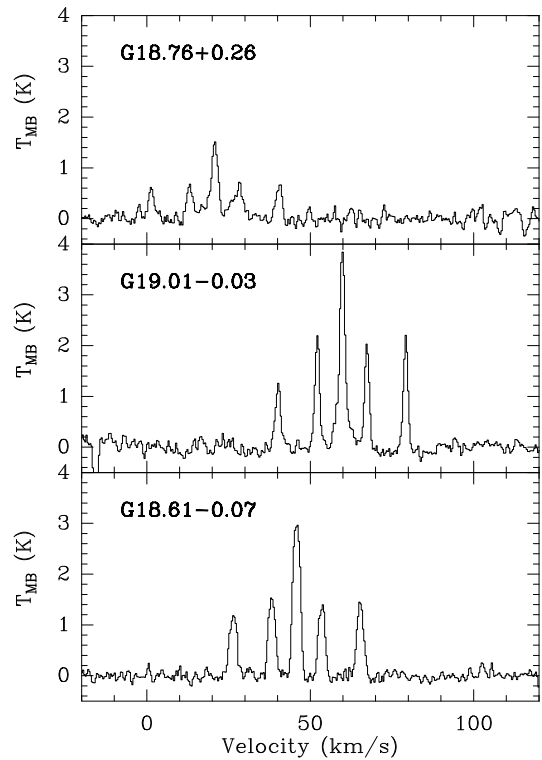}}
\caption{Example spectra in NH$_3$ (1,1) transition for three sources
located in the region shown in Fig.~\ref{fig-exmap}. G18.76+0.26 is
probably at a distance of $\sim$14~kpc, while G19.01-0.03 and G18.61-0.07
are probably at 4.3 and 3.7~kpc, respectively.}
\label{fig-ex-nh3}
\end{figure}

\begin{figure}
\centering
\includegraphics[width=8.5cm]{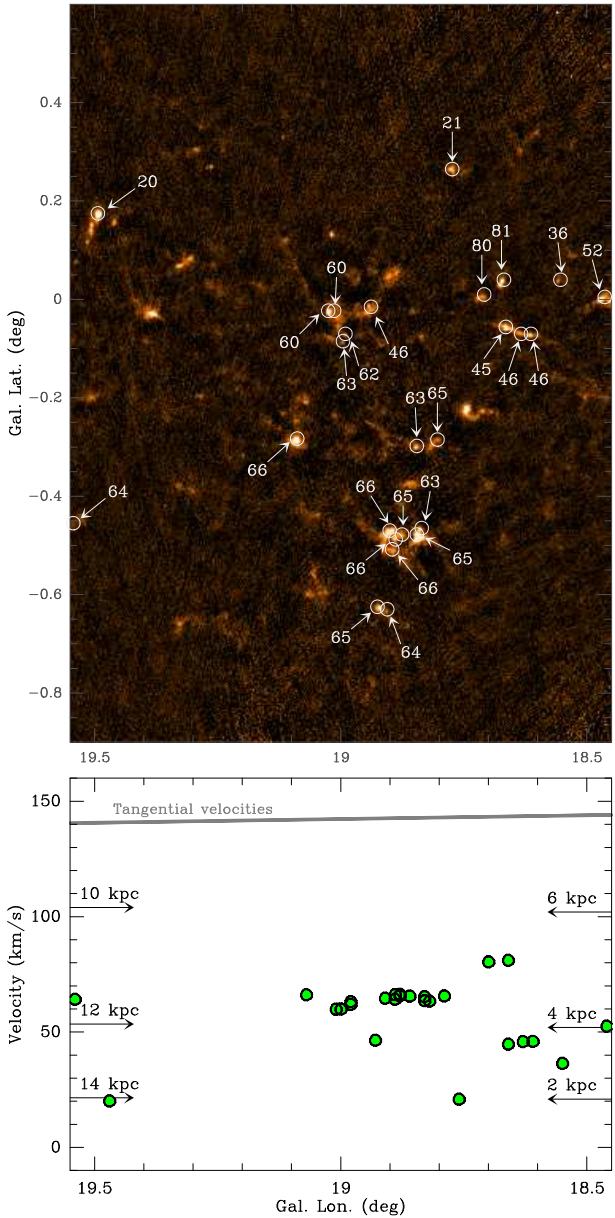}
\caption{{\em Top}: LSR velocities (in km~s$^{-1}$) derived from
NH$_3$ spectra are indicated next to the sources visible in the
ATLASGAL map. {\em Bottom}: the measured LSR velocities are plotted
against Galactic longitude. Kinematic distances derived using
the Galactic rotation model of \citet{brand-blitz} are indicated
on the right for the near distance solution, and on the left
for the far solution.
The thick line near 140~km~s$^{-1}$ shows the maximal expected
velocity, which corresponds to the tangential point for a circular
orbit with a radius $\sim$2.7~kpc around the GC.}
\label{fig-nh3-vel}
\end{figure}

Distance determination in the Galaxy is a difficult issue.
The kinematic distances can be obtained from molecular-line
follow-up surveys. We have started systematic observations of the compact
ATLASGAL sources in NH$_3$ with the Effelsberg 100~m telescope
(Wienen et al.~in prep.) for the northern part of the plane, and in many
molecular lines observed simultaneously with the MOPRA 22~m telescope
(Wyrowski et al.~in prep.) for the southern part.
The NH$_3$ lines were detected for nearly all targeted sources, showing
various line widths and shapes, and usually with only one
velocity component (see examples in Fig.~\ref{fig-ex-nh3}).
This is expected since NH$_3$ is a high-density tracer,
which is detectable only for the dense cores also seen with LABOCA.

Twenty-four of the total 278 sources observed so far in NH$_3$ are
located in the region shown in Fig.~\ref{fig-exmap}. The distribution
of measured V$_{\rm LSR}$ is shown in Fig.~\ref{fig-nh3-vel}.
Interestingly, the velocities are not spread over the entire permitted
range, but are mostly gathered in very few coherent groups. In addition,
these groups are also spatially coherent (see Fig.~\ref{fig-nh3-vel} top).
We can conclude that the ATLASGAL emission is dominated by a few bright
complexes at specific distances. In the $l = 19\degr$ region, 15 out of 24
ATLASGAL compact sources detected in NH$_3$ have a velocity of between
60 and 66~km~s$^{-1}$. This complex would be at a distance of 4 or 12~kpc
depending on whether the near or far solution for the kinematic distance
is correct. In addition to this clear velocity component, two sources are
at rather low velocities of around 20~km~s$^{-1}$, a few are between 35 and
50~km~s$^{-1}$, and two are at 80 and 81~km~s$^{-1}$.

\begin{figure}
\centering
\includegraphics[width=\columnwidth]{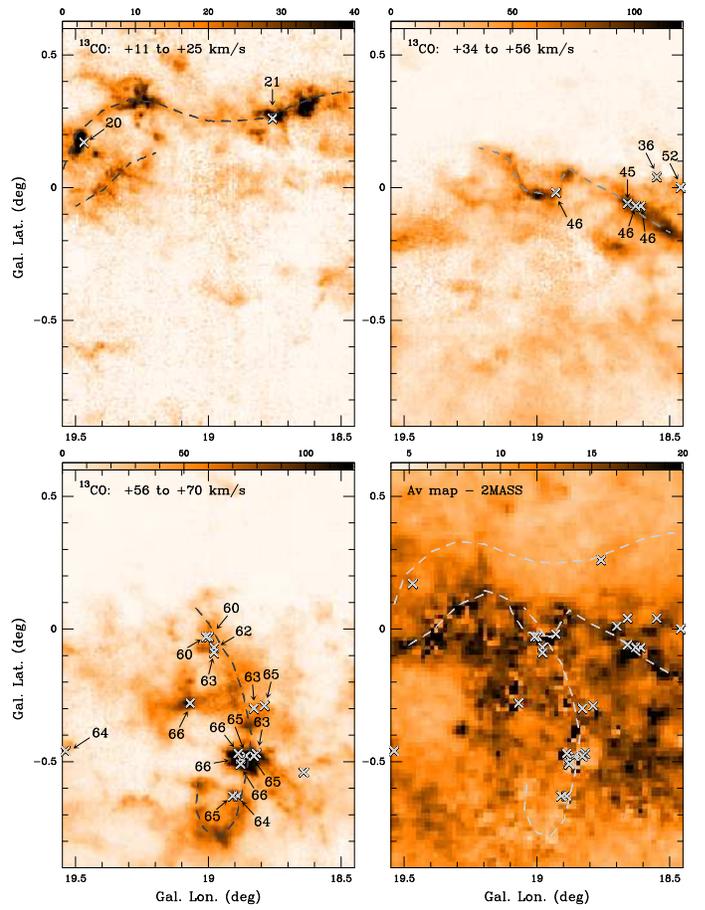}
\caption{Integrated $^{13}$CO(1--0) intensity over different ranges
of velocities, shown in the top-left corner of each panel, from
the GRS data. The color scale is shown on top of each panel, in units
of K~km~s$^{-1}$. LSR velocities of ATLASGAL sources derived from the
NH$_3$ spectra are indicated. {\em Bottom right}: extinction map
derived from the 2MASS catalog. The color scale corresponds to magnitudes
of visual extinction. In all panels, dashed lines outline structures
that seem connected together.}
\label{fig-grs-av}
\end{figure}

To study the spatial distribution of these velocity components in
greater detail, we used the $^{13}$CO(1--0) data from the GRS
to identify the molecular complexes associated with these velocities
(Fig.~\ref{fig-grs-av}). In addition, using an extinction map obtained
from the 2MASS catalog (Bontemps, priv. com.),
it is possible to recognize that some of these complexes
account for the detected extinction and are therefore most probably
at their near kinematic distance. This is the case for the 36 to
52~km~s$^{-1}$ complex, and for a small part of the $\sim$20~km~s$^{-1}$
clouds (dashed-line arc around (19.35, 0.05)
in Fig.~\ref{fig-grs-av}). The main complex at 65~km~s$^{-1}$ is more
complicated: parts of this complex are associated with peaks of
extinction, but other parts correspond to very low
extinction. Nevertheless, this might indicate that it
is also at the near distance. In addition,
this complex appears quite extended at 8\mic ~and
870\mic, which again favors the near distance solution.
The full extent of the $^{13}$CO emission, spanning about $0.9\degr$
in latitude, corresponds to 70~pc at a distance of 4.5~kpc,
typical of GMC dimensions.

\begin{figure}
\centering
\includegraphics[width=\columnwidth]{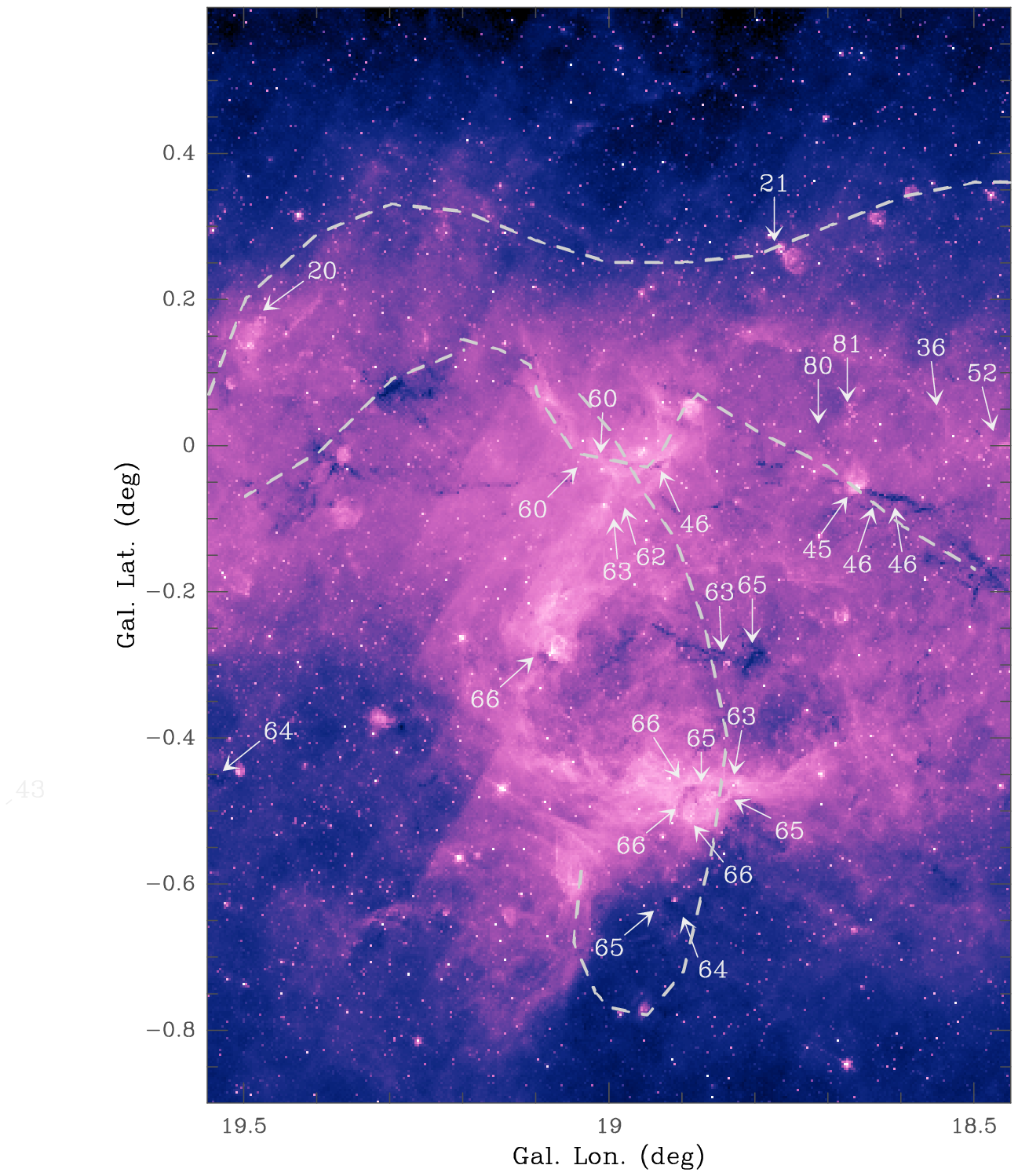}
\caption{8~$\mu$m map of the region shown in Fig.~\ref{fig-exmap} from the
GLIMPSE survey. LSR velocities derived from NH$_3$ data are indicated.}
\label{fig-glimpse}
\end{figure}

Finally, we note that a large part of the $\sim$20~km~s$^{-1}$
component is a string of compact CO clouds that are not associated
with any feature in the extinction map. They are therefore most
probably complexes at their far distance, i.e., on the order 14 kpc.
The source at $(l,b) = (18.76,0.26)$ coincides with a very compact
object in the GLIMPSE \citep{Benjamin2003} image
(Fig.~\ref{fig-glimpse}), which is also consistent
with a far distance solution.

Additional constraints can be obtained from absorption lines toward \hii
~regions \citep[e.g.,][]{ref-sewilo,ref-anderson}. \citet{ref-anderson}
presented seven \hii ~regions with measured distances determination
in the range $18.5\degr \, \leq \, l \, \leq \, 19.5\degr$: three
ultra-compact and four compact/classical regions. All three ultra-compact
and three out of the four compact regions have clear counterparts in
ATLASGAL (Fig.~\ref{fig-exmap}). Except for C18.95-0.02, which exhibits
several components in the $^{13}$CO spectrum, and C19.04-0.43, which we
did not observe in NH$_3$, there is good agreement between the LSR
velocities derived from radio recombination lines, from $^{13}$CO(1--0)
and from NH$_3$ transitions. Finally, results from, both,
\citet{ref-sewilo} and \citet{ref-anderson} generally agree with our
conclusions for the objects located in the test field, the +65~km~s$^{-1}$
complex being at the near distance and the +20~km~s$^{-1}$ clump G19.47+0.17
being at the far distance. However, in a few cases, the correct
interpretation of absorption lines remains difficult:
\citet{ref-sewilo} and \citet{ref-anderson} concluded with a far distance
solution for U18.66-0.06, but all three ATLASGAL sources that we observed
in NH$_3$ (with $v_{\rm LSR} \sim +45$~km~s$^{-1}$) follow an IRDC clearly
detected by GLIMPSE (Figs.~\ref{fig-glimpse} and \ref{fig-color}),
which favors a near distance.
Discussion of all individual sources goes beyond the scope of the
present paper, and will be addressed elsewhere.

\subsection{Mass estimates}

\begin{figure}[!tp]
\centering
\includegraphics[angle=270,width=8.5cm]{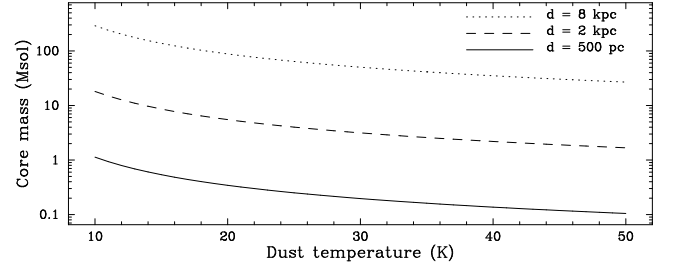}
\caption{Total gas plus dust masses in a core with 250~mJy integrated
flux density
at 870\mic ~(5-$\sigma$ detection limit) as a function of dust
temperature, for three typical distances: 500~pc (full line),
2~kpc (dashed line) and 8~kpc (dotted line).}
\label{fig-masses}
\end{figure}

Dust emission is generally optically thin in the submm continuum.
Therefore, following \citet{hildebrand}, the total mass in a core
is directly proportional to the total flux density $F_\nu$,
integrated over the source:
\begin{equation}
M \, = \, \frac{d^2 \, F_\nu \, R}{B_\nu(T_D) \, \kappa_\nu},
\end{equation}
where $d$ is the distance to the source, $R$ is the gas-to-dust mass ratio,
$B_\nu$ is the Planck function for a dust temperature $T_D$,
and $\kappa_\nu$ is the dust absorption coefficient.
The beam-averaged column density, which does not depend on the
distance, can also be computed with:
\begin{equation}
N_{H_2} \, = \, \frac{F_\nu \, R}{B_\nu(T_D) \, \Omega \, \kappa_\nu \, \mu
\, m_H},
\end{equation}
where $\Omega$ is the beam solid angle, $\mu$ is the mean molecular weight
of the interstellar medium, which we assume to be equal to 2.8, and
$m_H$ is the mass of an hydrogen atom.
Assuming a gas-to-dust mass ratio of 100, and
$\kappa_\nu \, = \, 1.85$~cm$^2$~g$^{-1}$ \citep[interpolated
to 870\mic ~from Table 1, Col.~5 of][]{ossenkopf+henning1994},
our survey is sensitive to cold cores with typical
masses in the range $\sim$100~M$_{\sun}$ ~at 8~kpc to below
1~M$_{\sun}$ ~at 500~pc (Fig.~\ref{fig-masses}).
The 5-$\sigma$ limit of 0.25~Jy/beam corresponds to column densities
in the range $3.6 \times 10^{21}$~cm$^{-2}$ for $T_D$~=~30~K
to $2.0 \times 10^{22}$~cm$^{-2}$ for $T_D$~=~10~K, equivalent
to visual extinctions of 4~mag to 21~mag, respectively
\citep[using the conversion factor from][]{ref-frerking}.

These computations assume that the emission at 870\mic ~is
optically thin. For the brightest source detected so far (Sgr~B2(N),
which has a peak flux density at 170~Jy/beam), we compute the
beam-averaged optical depth using:
\begin{equation}
\tau_{870} = - ln \left[ 1 - \frac{F_\nu}{\Omega B_\nu(T)} \right]
\end{equation}
This infers an optical depth of $\tau\leq0.31$ for a temperature
$T\geq25$~K. Therefore, the optically thin assumption should
be correct, even for the brightest sources.

\begin{table*}[!tp]
\caption{\label{tab-masses}Observed and derived parameters for the
24 sources located in the region shown in Fig.~\ref{fig-exmap} and
observed in NH$_3$ lines. When the distance ambiguity could be solved,
the preferred distance and mass are shown in boldface.}
\begin{tabular}{llllllllllllll}
\hline
\hline
Source & RA (J2000) & Dec (J2000) & Size & $v_{\rm LSR}$ &
$R_G$ & D$_{\rm near}$ & D$_{\rm far}$ &
F$_{\rm peak}$ & F$_{\rm int}$ & T$_{\rm kin}$ &
N$({\rm H_2})$ & M$_{\rm near}$ & M$_{\rm far}$ \\
& & & $''$ & km/s & kpc & kpc & kpc & Jy/b. &
Jy & K & $^{(1)}$ & M$_{\sun}$ & M$_{\sun}$ \\
\hline
 G18.55+0.04 & 18 24 37.85 & -12 45 14.8 & 36$\times$24 & 36.3 & 5.6 & 3.1 & 13.0 & 0.92 &  2.8 & 21.0 &  2.1 &  140 &  2380 \\
G18.61--0.07 & 18 25 08.42 & -12 45 21.6 & 38$\times$26 & 45.8 & 5.2 & {\bf 3.7} & 12.5 & 1.70 &  5.7 & 32.7$^{(2)}$ &  2.2 & {\bf  210} &  2490 \\
G18.63--0.07 & 18 25 09.84 & -12 44 07.5 & 53$\times$21 & 45.7 & 5.2 & {\bf 3.6} & 12.5 & 0.96 &  3.6 & 14.7 &  3.9 & {\bf  420} &  5010 \\
G18.65--0.06 & 18 25 10.63 & -12 42 24.1 & 43$\times$25 & 44.6 & 5.2 & {\bf 3.7} & 12.5 & 2.25 &  8.1 & 18.3 &  6.4 & {\bf  680} &  7940 \\
 G18.66+0.04 & 18 24 50.98 & -12 39 20.8 & 36$\times$24 & 80.9 & 4.0 & 5.2 & 10.9 & 2.10 &  6.3 & 22.4 &  4.5 &  780 &  3510 \\
 G18.70+0.00 & 18 25 02.38 & -12 38 12.6 & 46$\times$34 & 80.3 & 4.0 & 5.1 & 11.0 & 0.84 &  4.5 & 14.8 &  3.4 & 1030 &  4770 \\
 G18.76+0.26 & 18 24 13.27 & -12 27 44.6 & 59$\times$42 & 20.8 & 6.6 & 2.0 & {\bf 14.1} & 1.14 &  9.7 & 15.8 &  4.1 &  320 & {\bf 15090} \\
G18.79--0.29 & 18 26 15.53 & -12 41 36.2 & 40$\times$34 & 65.5 & 4.4 & {\bf 4.5} & 11.6 & 0.94 &  4.3 & 14.4 &  4.0 & {\bf  830} &  5360 \\
G18.83--0.30 & 18 26 23.70 & -12 39 38.9 & 37$\times$28 & 63.5 & 4.6 & {\bf 4.4} & 11.7 & 1.35 &  4.8 & -$^{(3)}$ &  5.3 & {\bf  800} &  5610 \\
G18.83--0.47 & 18 26 59.34 & -12 44 46.5 & 34$\times$32 & 63.3 & 4.5 & {\bf 4.5} & 11.6 & 1.50 &  5.7 & 18.9 &  4.1 & {\bf  680} &  4600 \\
G18.83--0.49 & 18 27 03.57 & -12 45 06.5 & 58$\times$25 & 65.2 & 4.5 & {\bf 4.5} & 11.6 & 1.34 &  6.7 & 20.8 &  3.2 & {\bf  720} &  4640 \\
G18.86--0.48 & 18 27 06.20 & -12 43 09.3 & 38$\times$23 & 65.4 & 4.4 & {\bf 4.6} & 11.5 & 0.73 &  2.1 & 18.8 &  2.0 & {\bf  270} &  1710 \\
G18.88--0.49 & 18 27 09.21 & -12 42 35.1 & 38$\times$25 & 65.5 & 4.5 & {\bf 4.5} & 11.6 & 1.11 &  3.5 & 22.4 &  2.4 & {\bf  340} &  2210 \\
G18.88--0.51 & 18 27 15.13 & -12 42 54.6 & 52$\times$29 & 66.2 & 4.4 & {\bf 4.6} & 11.5 & 0.62 &  3.2 & 23.6 &  1.2 & {\bf  290} &  1840 \\
G18.89--0.47 & 18 27 07.82 & -12 41 38.0 & 64$\times$34 & 66.1 & 4.4 & {\bf 4.6} & 11.5 & 3.39 & 25.4 & 25.9 &  5.9 & {\bf 2040} & 12840 \\
G18.89--0.64 & 18 27 43.46 & -12 45 48.5 & 28$\times$22 & 64.1 & 4.5 & {\bf 4.5} & 11.6 & 0.58 &  1.2 & 20.0 &  1.5 & {\bf  130} &   880 \\
G18.91--0.63 & 18 27 43.55 & -12 44 51.0 & 50$\times$43 & 64.5 & 4.5 & {\bf 4.5} & 11.6 & 1.51 & 11.1 & 17.1 &  4.8 & {\bf 1560} & 10340 \\
G18.93--0.01 & 18 25 32.35 & -12 26 41.4 & 27$\times$25 & 46.3 & 5.1 & 3.7 & 12.4 & 1.33 &  3.1 & 29.3 &  2.0 &  140 &  1530 \\
G18.98--0.07 & 18 25 51.06 & -12 25 42.5 & 26$\times$18 & 61.8 & 4.6 & {\bf 4.4} & 11.7 & 0.82 &  1.3 & 20.4 &  2.0 & {\bf  140} &   960 \\
G18.98--0.09 & 18 25 55.22 & -12 25 50.3 & 30$\times$19 & 63.0 & 4.5 & {\bf 4.4} & 11.6 & 0.64 &  1.3 & 20.7 &  1.5 & {\bf  130} &   900 \\
G18.99--0.04 & 18 25 45.44 & -12 23 54.0 & 38$\times$32 & 59.8 & 4.7 & {\bf 4.3} & 11.8 & 0.31 &  1.3 & 20.7 &  0.7 & {\bf  120} &   930 \\
G19.01--0.03 & 18 25 44.61 & -12 22 42.0 & 40$\times$34 & 59.8 & 4.7 & {\bf 4.3} & 11.8 & 2.18 & 10.1 & 19.5 &  5.7 & {\bf 1070} &  8000 \\
G19.08--0.29 & 18 26 48.77 & -12 26 20.7 & 39$\times$37 & 66.0 & 4.6 & {\bf 4.3} & 11.7 & 4.28 & 21.4 & 24.2 &  8.2 & {\bf 1700} & 12330 \\
 G19.47+0.17 & 18 25 54.50 & -11 52 34.4 & 33$\times$24 & 20.0 & 6.6 & 2.0 & {\bf 14.0} & 6.53 & 18.0 & 23.4 & 13.1 &  330 & {\bf 15510} \\

\hline
\end{tabular}

{\em Notes}: $^{(1)}$ Unit of N$({\rm H_2})$ is $10^{22}$ cm$^{-2}$ \\
 $^{(2)}$ This source was also observed in NH$_3$ by
\citet{sridharan+2005}, with a better signal-to-noise ratio. We
therefore use the kinetic temperature derived by this study rather
than our own data for this source.\\
$^{(3)}$ Only the (1,1) transition of NH$_3$ is detected at this
position, so that no kinetic temperature could be derived. A temperature
of 15~K has been assumed to compute masses and column density for this
source.
\end{table*}

Using the distances as estimated in Sect.~\ref{sec-distance}, we
calculated the masses of the clumps located around
$l = +19\degr$ (Fig.~\ref{fig-nh3-vel}); the results
are summarized in Table~\ref{tab-masses}.
In a few cases, the SExtractor program, used to extract compact
sources from the full survey (Sect.~\ref{sec-extraction}), does not
reliably extract sources that appear blended. Therefore, we fitted
2D-Gaussian profiles to the 24 sources observed in NH$_3$ to
measure their fluxes and sizes, special care being taken to separate
blended sources. The results of these fits were source positions, sizes
(minor and major axis FWHM), and both peak and integrated fluxes.
 
Table~\ref{tab-masses} provides: source names (based on Galactic
coordinates); equatorial coordinates and sizes, measured
from Gaussian fits on the 870\mic ~image;
LSR velocities measured with the NH$_3$(1,1) transition;
corresponding Galactocentric radius, as well as near and far
kinematic distances; peak and
integrated fluxes at 870\mic; and kinetic temperatures, computed
from the NH$_3$(1,1) and (2,2) measurements (Wienen et al.,
in prep.). Finally, masses for the near and far distances
are computed, using the integrated 870\mic ~fluxes and the
kinetic temperatures.

These computations assume that the 870\mic ~fluxes detected with
LABOCA originate only in thermal emission from the dust.
From spectral-line surveys, it is known that molecular lines might
contribute a significant amount of flux to the emission detected by
broadband bolometers. \citet{groesbeck1995} estimated from line surveys
in the 345~GHz atmospheric window that lines can contribute up to 60\% of
the flux. This extreme value was found for the Orion hot core, which is
peculiar due to a large contribution from wide SO$_2$ lines
\citep{schilke+1997}. The relative contribution from CO(3--2) lines
alone will only be of relevance in the case of
extreme outflow sources with little dust
continuum. From a quantitative analysis of the relative CO contribution
in the bright massive star-forming region G327.3-0.6, observed with the
APEX telescope in CO \citep{wyrowski+2006}, we found the contribution to
be insignificant close to the compact hot core but to increase
to 20\% for the bright, extended photon-dominated region to the north
of the hot core. In summary, hot molecular cores, strong outflow sources,
and bright photon-dominated regions will have a flux contribution from
line emission that is higher than the typical 15\% calibration uncertainty
(Sect.~\ref{sec-reduc1}) in the
continuum observations. Line follow-up observations in the mm/submm
with APEX of bright sources found with ATLASGAL will help to place
stronger constraints on the line flux contribution.

Another source of contamination could be free-free emission, when
an \hii ~region has already been formed. At 345~GHz the contribution from
free-free should however be low compared to the thermal emission from
dust. As an example, \citet{motte+2003} estimated the contribution
from free-free emission at 1.3~mm in the compact fragments found within
the W43 massive star-forming region, which harbors a giant \hii ~region.
In most cases, the free-free emission accounts for $\leq$40\% at
1.3~mm, one extreme case being where this fraction reaches 70\%.
Assuming a flat spectrum for free-free (optically thin case), and
$F_{\nu} \propto \nu^{3.5}$ for the dust emission (i.e., a dust emissivity
of $\beta \, = \, 1.5$), scaling from 240~GHz to 345~GHz implies that 20\%
of the flux originates from free-free emission at the LABOCA
wavelength. In conclusion, the contribution from free-free should be almost
negligible, and even in the most extreme cases should not exceed 20\%.

\subsection{Nature of the sources}

\begin{figure}
\centering
\includegraphics[width=9cm]{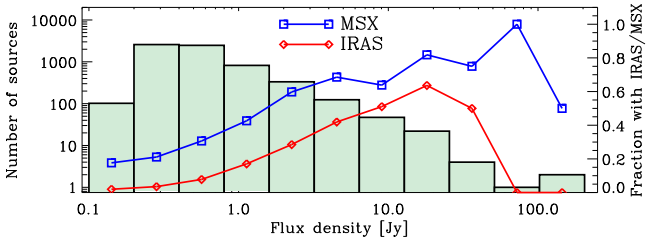}
\caption{Distribution of peak fluxes for the compact sources extracted
in the data observed in 2007, shown in logarithmic scale.
Fractions of sources associated with IRAS or MSX sources within a 30$''$
search radius are indicated, using the scale shown in the right hand
side.}
\label{fig-fluxes}
\end{figure}

Since dust emission in the submm is optically thin, the LABOCA
observations are sensitive to all kinds of objects containing dust,
which can range from cold, dense clumps within clouds to more evolved
proto-stars that are centrally heated. We searched systematically for
infrared counterparts to our preliminary catalog of 6000 sources in
the IRAS Point Source Catalog \citep{Beichman1988} and in the Midcourse
Space Experiment \citep[MSX,][]{Price2001} data, using a search radius of
30$''$ (more details will be given in Contreras et al.~in prep.).
Only $\sim$10\% of the submm compact sources can be associated with
an IRAS point source, and $\sim$32\% have a possible MSX counterpart
(Fig.~\ref{fig-fluxes}). In contrast, two thirds do not have a mid- or
far-infrared counterpart in the IRAS/MSX catalogs, and could correspond
to the coldest stage of star formation, prior to the birth of a massive
proto-star. This fraction is similar to the results found by
\citet{motte+2007} for the Cygnus-X region. However, with the superior
sensitivity of the Spitzer GLIMPSE and MIPSGAL \citep{Carey2005}
surveys, weaker mid-IR sources are found to be associated with many
of the MSX-dark, compact submm-clumps. Sources with IRAS associations
generally correspond to later stages of star formation, e.g., hot molecular
cores, compact \hii ~regions, or young embedded star clusters.

One should note that in some bright, complex star-forming regions,
because of crowding and blending, even strong infrared sources do not
fulfill the point-source selection criteria and, consequently, are not
included in the IRAS/MSX catalogs. An extreme example is seen in
the histogram of Fig.~\ref{fig-fluxes}, where the brightest source
in the survey, SgrB2(N), is counted as a source without IRAS/MSX
association.

\begin{figure}
\centering
\includegraphics[width=8.5cm]{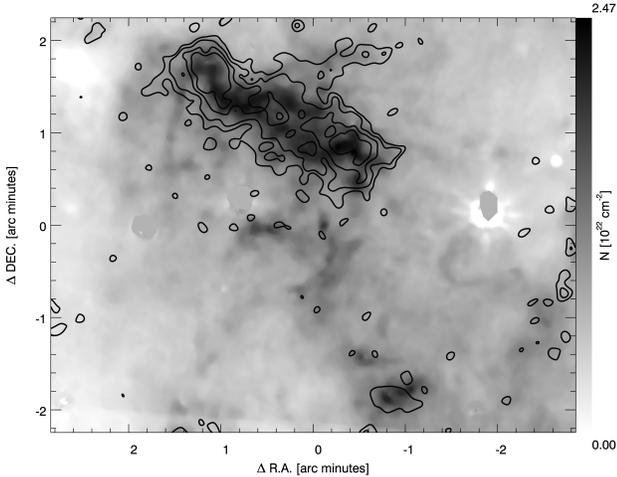}
\caption{Column density map of the IRDC G19.30+0.07, contained in the $l =
19^\circ$ demonstration field. The map is derived from 8\mic ~extinction
values measured in the corresponding Spitzer/GLIMPSE map. The ATLASGAL
emission at 870~$\mu$m is overlaid in contours. The map center is at
$\alpha \, = \, 18^h\,25^m\,53.94^s$, $\delta \, = \, -12\degr\,05'43.9''$
(J2000), or $l \, = \, 19.2771\degr$, $b \, = \, +0.0703\degr$.}
\label{fig-IRDC}
\end{figure}

High-contrast IRDCs are easily identified by ATLASGAL. These clouds can
have various shapes, from compact cores to filamentary morphologies of
1--5 arcmin in size. As an example, our demonstration field contains the
already well-investigated IRDC G19.30+0.07 \citep{carey+1998}. In
Fig.~\ref{fig-IRDC}, we show a column-density map of this IRDC, based on
an extinction map derived from GLIMPSE
8\mic ~data. The $R_{\rm V} = 5.5$ extinction law of
\citet{weingartner} has been used. Further details of this approach
are described in a forthcoming publication \citep{ref-vasy}. Overlaid as
contours are the ATLASGAL 870\mic ~emission data. There is good agreement
between the morphologies observed in both datasets, especially at the
high column-density center of the IRDC. With an adopted distance of 2.38 kpc,
roughly 260 M$_{\sun}$ are collected in this clump. IRDCs are considered
to be a good hunting ground for the earliest phases of star formation, and
ATLASGAL is expected to reveal a considerable fraction of them to be sub-mm
emission sources.

Some sources seen in the submm may also correspond to {\em low-mass}
star-forming sites located at small distances.
However, our survey is sensitive to 1~M$_{\sun}$ dense clumps up to
a distance of $\sim$1~kpc only (Eq.~(1)). In addition, nearby
regions of star formation are likely to be located away from the Galactic
Plane, outside the area covered by ATLASGAL. Therefore, low-mass
star-forming clumps are not expected to be a dominant population
in our data.

A small fraction of sources detected in the submm may not be related to
star formation. These sources include late evolved stars with circumstellar
envelopes, planetary nebulae, the nuclei of external galaxies, and quasars.
However, we estimate that the contamination due to these classes of objects
is negligible. For instance, following the method outlined by
\citet[][Appendix]{anglada1998}, the number of background sources
per arcmin$^2$ with flux densities at 5~GHz above S is:
\begin{equation}
N(arcmin^{-2}) = 0.011 \times S(mJy)^{-0.75}
\end{equation}
Assuming that the flux density scales with frequency as $S \propto
\nu^{-0.7}$, one then obtains, for $\nu \, = \, 345$~GHz, 0.10 sources
above 150~mJy per square degree, or about 10 background sources in the
95~deg$^2$ area discussed here. The extrapolation from 5~GHz to 345~GHz
is quite uncertain, but this example illustrates nevertheless that the
number of these background sources must be very low.

Another estimate can be derived from deep-field surveys in the
submm. Integrating the differential source counts from the
SCUBA Half-Degree Extragalactic Survey at 850\mic
~\citep[SHADES,][their Eq.~12 and Table~7]{ref-shades}, yields
$1 \times 10^{-3}$ sources per square degree with flux densities
above 150~mJy, or 0.1 sources above this limit in our surveyed
area. Thus, we do not expect to detect any high-redshift
object in our data.

The situation for late evolved stars may be more complex. As an example,
the red supergiant VY~CMa, which is one of the brightest sources at IR
wavelengths and is at a distance of $\sim$1.5~kpc,
has a flux density of 1.5~Jy at 870\mic, as measured with LABOCA.
Therefore, this object is detectable out to a distance of 3.5~kpc
with our survey sensitivity. Less extreme or more distant
objects of similar nature will most likely fall below our detection
limits, but we cannot exclude having a small fraction of supergiants
and extreme late evolved stars in our data. It will be possible to
identify such objects from their colors at other wavelengths,
e.g., in the Spitzer surveys \citep{robitaille2008}.

\section{\label{sec-persp}Perspectives}

\subsection{An unbiased survey}

This survey will provide the first unbiased database of Galactic
sources at submm wavelengths, covering the inner $120\degr$ of the
Galaxy in a systematic fashion. Since dust emission in the submm is
optically thin, it is directly proportional to
the mass of material (assuming a constant temperature)
and leads to the detection of all proto-stellar
objects or pre-stellar clumps and cores. With a 5-$\sigma$ detection
limit of 250~mJy/beam, this survey will be complete for proto-stellar cores
down to $\sim$10~M$_{\sun}$ at 2~kpc, or $\sim$50~M$_{\sun}$
at 5~kpc (Fig.~\ref{fig-masses}).
In addition, the data will allow the investigation of the structure of
molecular complexes on various scales, from their
lower-intensity filaments to higher-density star-forming regions.

The systematic nature of this survey will also allow a complete census of
extreme sources in the inner Galaxy, such as the high-mass star-forming
regions W49 or W51. With our sensitivity, we should detect, at a
10-$\sigma$ level, sources on the far side of the Galaxy, 25~kpc away,
of masses as low as about 1500~M$_{\sun}$.

The unbiased nature of this database is critically important in compiling
robust samples of objects for all types of follow-up projects based
on a reliable statistical analysis. This is especially
true in the context of ALMA (see also Sect.~\ref{sec-legacy}
below), which will explore exactly the same part of the sky as APEX
in the (sub)mm range.

\subsection{Evolutionary stages of high-mass star-formation}

This database will contain a large number of high-mass clouds with submm
emission not associated with developed star formation (as traced in the
infrared surveys), which are expected to correspond to the earliest phases
during the formation of the richest clusters of the Galaxy. Since
high-mass proto-stars are rare \citep{motte+2007}, such large-scale surveys
are required to obtain large samples of objects corresponding to the
different stages of massive-star formation. Thus, the data will allow us to
determine a basic evolutionary sequence for high-mass star formation.

Another topic that can be addressed with these data is the
importance of triggered star formation, which can be assessed
by identifying the second generation stars at the edges of \hii
~regions \citep[e.g.,][]{ref-deharveng} and supernova remnants.
In the context of high-mass star formation in particular, the importance
of triggering compared to spontaneous star formation remains
an open question.

\subsection{Distances and Galactic structure}

Dust emission relates the star-forming regions seen in the IR/radio to the
molecular clouds seen in CO, e.g., in the Galactic Ring Survey, by tracing
the highest column-density features: compact sources and filaments.
With the help of both extinction maps and absorption lines toward \hii
~regions \citep{ref-sewilo}, the near/far ambiguity of kinematic distances
can be solved in deriving the proper distances for a large fraction of
ATLASGAL sources (Sect.~\ref{sec-distance}). This will therefore help to
improve our 3D view of the Galaxy. These data will also allow us to
discover new coherent star-formation complexes up to distances of a few kpc.
Finally, these data will allow us to compare the star-formation processes
at different locations, in the spiral arms and inter-arm regions, and
between the different arms.

\subsection{\label{sec-database}Ancillary data}

\begin{figure}[tp]
\hspace{13mm}
\includegraphics[width=7.5cm]{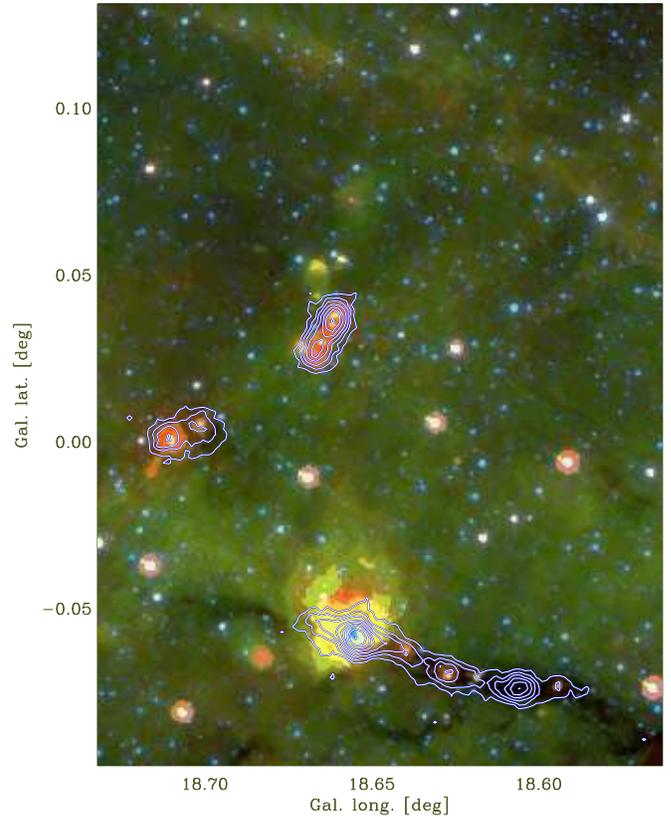}
\vspace{6mm}
\caption{Three-color image from Spitzer surveys: GLIMPSE 4.5\mic
~in blue (mostly tracing stars) and 8\mic ~in green (mostly PAH
emission), MIPSGAL 24\mic ~in red, with 870\mic ~emission overlaid
in contours.}
\label{fig-color}
\end{figure}

The region that we plan to map covers the same area as the GLIMPSE and
MIPSGAL surveys (e.g., Fig.~\ref{fig-color}). It will also be mapped with
PACS and SPIRE for the approved Hi-GAL (the Herschel infrared Galactic
Plane Survey; Molinari et al.) open-time Herschel key project. The
ATLASGAL data will thus be cross-correlated with similar large-scale
Galactic surveys at other wavelengths, in the near- to far-infrared range,
for example with the $J, H, K$ UKIDSS Galactic Plane Survey \citep{Lucas2007},
providing spectral energy distributions from $\sim$1 to 870$\,\mu$m. The
LABOCA data at 870\mic ~are crucial to estimating the total amount of dust
along the line of sight. The combination of data from APEX, Spitzer, and
Herschel will provide an extraordinary new database of the Galactic Plane
from the near-IR to the millimeter, at a spatial resolution $\la$30$''$.

We will also correlate our data with existing radio surveys in the
centimeter range: the NRAO VLA Sky Survey \citep[NVSS,][]{Condon1998}, the
Multi-Array Galactic Plane Imaging Survey \citep[MAGPIS,][]{Helfand2006},
the Sydney University Molonglo Sky Survey \citep[SUMSS,][]{ref-sumss},
and the Co-Ordinated Radio `N' Infrared Survey for High-mass star formation
(CORNISH)\footnote{\url{http://www.ast.leeds.ac.uk/Cornish/index.html}}.
It will be of particular interest to compare with surveys of the radio
continuum and maser emission that will be possible with the Expanded
Very Large Array and the Australia Telescope Compact Array in the near
future \citep{Menten2007}.
Existing or upcoming molecular-line databases include those resulting
from the multi-beam CH$_3$OH maser Galactic Plane survey \citep{Green2007}
and from surveys, e.g., in the CS or $^{13}$CO molecules
\citep{Bronfman1996, Urquhart2007}.
Combining ATLASGAL with the IR to cm data will allow us
to establish a solid picture of the star-formation evolutionary
stages and their associated timescales.

Several bolometer arrays are expected to be installed at APEX in the
near future, including the Submillimetre APEX Bolometer Camera (SABOCA),
a 37-element array observing at 350\mic, with a beam of $7\farcs7$. Our
team plans to complement the 870\mic ~data with targeted observations at
350\mic, covering limited areas of a few arcmin extent, to constrain the
dust emissivity and resolve bright sources in crowded regions. The 870\mic
~survey might also be extended in surface area (toward the outer Galactic
Plane), or to greater depth for a small part of the Galactic Plane
(20~mJy/beam rms or less).

A similar survey of the northern Galactic Plane ($l \, \geq \, -10\degr$)
has started with the BOLOCAM instrument at the Caltech Submillimeter
Observatory (CSO), at a wavelength of 1.1~mm \citep{bolocam-gps}, with a
27$''$ resolution.
The northern Galactic Plane will also be mapped with the SCUBA-2
instrument at 450 and 850\mic, with resolutions of 8$''$ and 15$''$,
respectively. In particular, as part of the SCUBA-2 ``All-Sky''
Survey \citep[SASSY,][]{ref-sassy}, a $10\degr$ wide strip covering
$\vert b \vert \, \leq \, 5\degr$ over the part of the Galactic Plane
accessible from Hawaii will be observed at 850\mic ~with a one-$\sigma$
sensitivity of 30~mJy/beam, comparable to the ATLASGAL survey.
In addition, the JCMT Galactic Plane Survey plans to cover much deeper
the area $+10\degr \leq l \leq +65\degr$,
$\vert b \vert \leq 1\degr$, at both 450 and 850\mic.
Combining data from these large-scale surveys will
provide data at three different wavelengths in the Rayleigh-Jeans
part of the dust emission spectrum, thus allowing us to study,
among others, the dust emissivity, related to the composition of
the dust.

\subsection{\label{sec-legacy}Legacy value}

The raw data will be available to outside users through the ESO
archive\footnote{\url{http://archive.eso.org/wdb/wdb/eso/apex/form}}
one year after the end of each observing period. We also plan to
make the reduced combined maps available to the community;
in addition, a compact-source catalog (including
associations with surveys at other wavelengths) and a catalog of
extended objects will be published upon completion. These will be
available from a public web
site\footnote{\url{http://www.mpifr-bonn.mpg.de/div/atlasgal/ }}.
This project
will thus have a high legacy value, providing well characterized
samples of targets to address in detail many questions of
astrophysics.
Calibrated maps built from the data observed in 2007 and
discussed in the present paper are expected to be available online,
both on the ESO archive and the ATLASGAL public website,
by Summer 2009.

The observations that we have planned should be completed by the end
of 2009, i.e. when the Herschel observatory is expected to deliver its
first science data. Our
team will be involved in the investigation of systematic associations
between the
ATLASGAL sources and Herschel products. In particular, members of
the consortium are leading the key projects HOBYS (the Herschel
imaging survey of OB Young Stellar objects; Motte, Zavagno, Bontemps
et al.) and Hi-GAL.

The first ALMA antennas should also be available for early science
soon after completion of our survey. This will be
the perfect time for conducting follow-ups of carefully selected targets
at an unprecedented spatial resolution. The ATLASGAL catalog will provide
ALMA with more than 10$^4$ compact sources, including hot cores, high-mass
proto-stellar objects, and cold pre-stellar cores. All sorts of molecular
lines will be available to study the chemistry of the sources, or to search
for signatures of infall or outflows for example.
Thus, ATLASGAL is expected to trigger numerous follow-up
studies to characterize the detected sources in terms of their excitation
and chemistry. ATLASGAL will also provide the James Webb Space
Telescope (JWST) with samples of well characterized targets,
and these future facilities will allow one
to study the inner workings of the star-formation process on small
scales. This survey is therefore an essential pathfinder
for Herschel, ALMA, and the JWST.

In addition to the new findings in the realm of high-mass star formation,
the survey will tell us on a larger scale details about the structure and
mass distribution in our Galaxy and help us to connect our knowledge of
local star formation with star formation in external galaxies, where
high-mass stars are the only ones that we see.

\begin{acknowledgements}
We thank the APEX staff for their help during the observations, and for
their continuous support with LABOCA operations.
This research has made use of the SIMBAD database, and of the Aladin Sky Atlas,
both operated at CDS, Strasbourg, France.
H.B. acknowledges financial support by the Emmy-Noether-Program of the Deutsche
Forschungsgemeinschaft (DFG, grant BE2578).
L.B., G.G., and D.M. acknowledge support from Chilean Center of Excellence in
Astrophysics and Associated Technologies (PFB 06), and from Chilean Center
for Astrophysics FONDAP 15010003.
\end{acknowledgements}

\bibliographystyle{aa}
\bibliography{ATLASGAL}

\begin{thebibliography}{81}
\expandafter\ifx\csname natexlab\endcsname\relax\def\natexlab#1{#1}\fi

\bibitem[{{Adams} {et~al.}(1987){Adams}, {Lada}, \& {Shu}}]{adams+1987}
{Adams}, F.~C., {Lada}, C.~J., \& {Shu}, F.~H. 1987, \apj, 312, 788

\bibitem[{{Anderson} \& {Bania}(2009)}]{ref-anderson}
{Anderson}, L.~D. \& {Bania}, T.~M. 2009, \apj, 690, 706

\bibitem[{{Andr\'e} {et~al.}(2000){Andr\'e}, {Ward-Thompson}, \&
  {Barsony}}]{andre-pp4}
{Andr\'e}, P., {Ward-Thompson}, D., \& {Barsony}, M. 2000, Protostars and
  Planets IV, 59

\bibitem[{{Anglada} {et~al.}(1998){Anglada}, {Villuendas}, {Estalella},
  {Beltr{\'a}n}, {Rodr{\'{\i}}guez}, {Torrelles}, \& {Curiel}}]{anglada1998}
{Anglada}, G., {Villuendas}, E., {Estalella}, R., {et~al.} 1998, \aj, 116, 2953

\bibitem[{{Beichman} {et~al.}(1988){Beichman}, {Neugebauer}, {Habing}, {Clegg},
  \& {Chester}}]{Beichman1988}
{Beichman}, C.~A., {Neugebauer}, G., {Habing}, H.~J., {Clegg}, P.~E., \&
  {Chester}, T.~J., eds. 1988, {Infrared astronomical satellite (IRAS) catalogs
  and atlases. Volume 1: Explanatory supplement}, Vol.~1

\bibitem[{{Benjamin} {et~al.}(2003){Benjamin}, {Churchwell}, {Babler}, {Bania},
  {Clemens}, {Cohen}, {Dickey}, {Indebetouw}, {Jackson}, {Kobulnicky},
  {Lazarian}, {Marston}, {Mathis}, {Meade}, {Seager}, {Stolovy}, {Watson},
  {Whitney}, {Wolff}, \& {Wolfire}}]{Benjamin2003}
{Benjamin}, R.~A., {Churchwell}, E., {Babler}, B.~L., {et~al.} 2003, \pasp,
  115, 953

\bibitem[{{Bertin} \& {Arnouts}(1996)}]{ref-sextractor}
{Bertin}, E. \& {Arnouts}, S. 1996, \aaps, 117, 393

\bibitem[{{Beuther} {et~al.}(2007){Beuther}, {Churchwell}, {McKee}, \&
  {Tan}}]{beuther+2007}
{Beuther}, H., {Churchwell}, E.~B., {McKee}, C.~F., \& {Tan}, J.~C. 2007, in
  Protostars and Planets V, ed. B.~{Reipurth}, D.~{Jewitt}, \& K.~{Keil},
  165--180

\bibitem[{{Beuther} {et~al.}(2002){Beuther}, {Walsh}, {Schilke}, {Sridharan},
  {Menten}, \& {Wyrowski}}]{beuther-masers}
{Beuther}, H., {Walsh}, A., {Schilke}, P., {et~al.} 2002, \aap, 390, 289

\bibitem[{{Bonnell} {et~al.}(2004){Bonnell}, {Vine}, \& {Bate}}]{bonnell2004}
{Bonnell}, I.~A., {Vine}, S.~G., \& {Bate}, M.~R. 2004, \mnras, 349, 735

\bibitem[{{Brand} \& {Blitz}(1993)}]{brand-blitz}
{Brand}, J. \& {Blitz}, L. 1993, \aap, 275, 67

\bibitem[{{Bronfman} {et~al.}(2000){Bronfman}, {Casassus}, {May}, \&
  {Nyman}}]{bronfman2000}
{Bronfman}, L., {Casassus}, S., {May}, J., \& {Nyman}, L.-{\AA}. 2000, \aap,
  358, 521

\bibitem[{{Bronfman} {et~al.}(1996){Bronfman}, {Nyman}, \&
  {May}}]{Bronfman1996}
{Bronfman}, L., {Nyman}, L.-A., \& {May}, J. 1996, \aaps, 115, 81

\bibitem[{{Carey} {et~al.}(1998){Carey}, {Clark}, {Egan}, {Price}, {Shipman},
  \& {Kuchar}}]{carey+1998}
{Carey}, S.~J., {Clark}, F.~O., {Egan}, M.~P., {et~al.} 1998, \apj, 508, 721

\bibitem[{{Carey} {et~al.}(2005){Carey}, {Noriega-Crespo}, {Price}, {Padgett},
  {Kraemer}, {Indebetouw}, {Mizuno}, {Ali}, {Berriman}, {Boulanger}, {Cutri},
  {Ingalls}, {Kuchar}, {Latter}, {Marleau}, {Miville-Deschenes}, {Molinari},
  {Rebull}, \& {Testi}}]{Carey2005}
{Carey}, S.~J., {Noriega-Crespo}, A., {Price}, S.~D., {et~al.} 2005, in
  Bulletin of the American Astronomical Society, Vol.~37, 1252

\bibitem[{{Condon} {et~al.}(1998){Condon}, {Cotton}, {Greisen}, {Yin},
  {Perley}, {Taylor}, \& {Broderick}}]{Condon1998}
{Condon}, J.~J., {Cotton}, W.~D., {Greisen}, E.~W., {et~al.} 1998, \aj, 115,
  1693

\bibitem[{{Coppin} {et~al.}(2006){Coppin}, {Chapin}, {Mortier}, {Scott},
  {Borys}, {Dunlop}, {Halpern}, {Hughes}, {Pope}, {Scott}, {Serjeant}, {Wagg},
  {Alexander}, {Almaini}, {Aretxaga}, {Babbedge}, {Best}, {Blain}, {Chapman},
  {Clements}, {Crawford}, {Dunne}, {Eales}, {Edge}, {Farrah}, {Gazta{\~n}aga},
  {Gear}, {Granato}, {Greve}, {Fox}, {Ivison}, {Jarvis}, {Jenness}, {Lacey},
  {Lepage}, {Mann}, {Marsden}, {Martinez-Sansigre}, {Oliver}, {Page},
  {Peacock}, {Pearson}, {Percival}, {Priddey}, {Rawlings}, {Rowan-Robinson},
  {Savage}, {Seigar}, {Sekiguchi}, {Silva}, {Simpson}, {Smail}, {Stevens},
  {Takagi}, {Vaccari}, {van Kampen}, \& {Willott}}]{ref-shades}
{Coppin}, K., {Chapin}, E.~L., {Mortier}, A.~M.~J., {et~al.} 2006, \mnras, 372,
  1621

\bibitem[{{Deharveng} {et~al.}(2005){Deharveng}, {Zavagno}, \&
  {Caplan}}]{ref-deharveng}
{Deharveng}, L., {Zavagno}, A., \& {Caplan}, J. 2005, \aap, 433, 565

\bibitem[{{Deharveng} {et~al.}(2009){Deharveng}, {Zavagno}, {Schuller},
  {Caplan}, {Pomar{\`e}s}, \& {De Breuck}}]{ref-rcw120}
{Deharveng}, L., {Zavagno}, A., {Schuller}, F., {et~al.} 2009, \aap, 496, 177

\bibitem[{{Di Francesco} {et~al.}(2008){Di Francesco}, {Johnstone}, {Kirk},
  {MacKenzie}, \& {Ledwosinska}}]{scuba-legacy}
{Di Francesco}, J., {Johnstone}, D., {Kirk}, H., {MacKenzie}, T., \&
  {Ledwosinska}, E. 2008, \apjs, 175, 277

\bibitem[{{Dowell} {et~al.}(2003){Dowell}, {Allen}, {Babu}, {Freund},
  {Gardner}, {Groseth}, {Jhabvala}, {Kovacs}, {Lis}, {Moseley}, {Phillips},
  {Silverberg}, {Voellmer}, \& {Yoshida}}]{ref-sharc2}
{Dowell}, C.~D., {Allen}, C.~A., {Babu}, R.~S., {et~al.} 2003, in Society of
  Photo-Optical Instrumentation Engineers (SPIE) Conference Series, ed. T.~G.
  {Phillips} \& J.~{Zmuidzinas}, Vol. 4855, 73--87

\bibitem[{{Drosback} {et~al.}(2008){Drosback}, {Aguirre}, {Bally}, {Bradley},
  {Chamberlin}, {Cyganowski}, {Evans}, {Ginsburg}, {Glenn}, {Harvey},
  {Nordhaus}, {Rosolowsky}, {Stringfellow}, {Vaillancourt}, {Walawender}, \&
  {Williams}}]{bolocam-gps}
{Drosback}, M.~M., {Aguirre}, J., {Bally}, J., {et~al.} 2008, in American
  Astronomical Society Meeting Abstracts, Vol. 212, 96.01

\bibitem[{{Enoch} {et~al.}(2006){Enoch}, {Young}, {Glenn}, {Evans}, {Golwala},
  {Sargent}, {Harvey}, {Aguirre}, {Goldin}, {Haig}, {Huard}, {Lange},
  {Laurent}, {Maloney}, {Mauskopf}, {Rossinot}, \& {Sayers}}]{enoch+2006}
{Enoch}, M.~L., {Young}, K.~E., {Glenn}, J., {et~al.} 2006, \apj, 638, 293

\bibitem[{{Evans} {et~al.}(2003){Evans}, {Allen}, {Blake}, {Boogert}, {Bourke},
  {Harvey}, {Kessler}, {Koerner}, {Lee}, {Mundy}, {Myers}, {Padgett},
  {Pontoppidan}, {Sargent}, {Stapelfeldt}, {van Dishoeck}, {Young}, \&
  {Young}}]{ref-c2d}
{Evans}, II, N.~J., {Allen}, L.~E., {Blake}, G.~A., {et~al.} 2003, \pasp, 115,
  965

\bibitem[{{Fa{\'u}ndez} {et~al.}(2004){Fa{\'u}ndez}, {Bronfman}, {Garay},
  {Chini}, {Nyman}, \& {May}}]{faundez+2004}
{Fa{\'u}ndez}, S., {Bronfman}, L., {Garay}, G., {et~al.} 2004, \aap, 426, 97

\bibitem[{{Frerking} {et~al.}(1982){Frerking}, {Langer}, \&
  {Wilson}}]{ref-frerking}
{Frerking}, M.~A., {Langer}, W.~D., \& {Wilson}, R.~W. 1982, \apj, 262, 590

\bibitem[{{Gahm} {et~al.}(2002){Gahm}, {Lehtinen}, {Carlqvist}, {Harju},
  {Juvela}, \& {Mattila}}]{gahm2002}
{Gahm}, G.~F., {Lehtinen}, K., {Carlqvist}, P., {et~al.} 2002, \aap, 389, 577

\bibitem[{{Garay} {et~al.}(2004){Garay}, {Fa{\'u}ndez}, {Mardones}, {Bronfman},
  {Chini}, \& {Nyman}}]{garay+2004}
{Garay}, G., {Fa{\'u}ndez}, S., {Mardones}, D., {et~al.} 2004, \apj, 610, 313

\bibitem[{{Green} {et~al.}(2007){Green}, {Cohen}, {Caswell}, {Fuller},
  {Brooks}, {Burton}, {Chrysostomou}, {Diamond}, {Ellingsen}, {Gray}, {Hoare},
  {Masheder}, {McClure-Griffiths}, {Pestalozzi}, {Phillips}, {Quinn},
  {Thompson}, {Voronkov}, {Walsh}, {Ward-Thompson}, {Wong-McSweeney}, {Yates},
  \& {Cox}}]{Green2007}
{Green}, J.~A., {Cohen}, R.~J., {Caswell}, J.~L., {et~al.} 2007, in IAU
  Symposium, Vol. 242, IAU Symposium, 218--222

\bibitem[{{Groesbeck}(1995)}]{groesbeck1995}
{Groesbeck}, T.~D. 1995, PhD thesis, California Institute of Technology.

\bibitem[{{G{\"u}sten} {et~al.}(2006{\natexlab{a}}){G{\"u}sten}, {Booth},
  {Cesarsky}, {Menten}, {Agurto}, {Anciaux}, {Azagra}, {Belitsky}, {Belloche},
  {Bergman}, {De Breuck}, {Comito}, {Dumke}, {Duran}, {Esch}, {Fluxa}, {Greve},
  {Hafok}, {H{\"a}upl}, {Helldner}, {Henseler}, {Heyminck}, {Johansson},
  {Kasemann}, {Klein}, {Korn}, {Kreysa}, {Kurz}, {Lapkin}, {Leurini}, {Lis},
  {Lundgren}, {Mac-Auliffe}, {Martinez}, {Melnick}, {Morris}, {Muders},
  {Nyman}, {Olberg}, {Olivares}, {Pantaleev}, {Patel}, {Pausch}, {Philipp},
  {Philipps}, {Sridharan}, {Polehampton}, {Reveret}, {Risacher}, {Roa},
  {Sauer}, {Schilke}, {Santana}, {Schneider}, {Sepulveda}, {Siringo},
  {Spyromilio}, {Stenvers}, {van der Tak}, {Torres}, {Vanzi}, {Vassilev},
  {Weiss}, {Willmeroth}, {Wunsch}, \& {Wyrowski}}]{guesten+2006a}
{G{\"u}sten}, R., {Booth}, R.~S., {Cesarsky}, C., {et~al.} 2006{\natexlab{a}},
  in SPIE Conference, Vol. 6267, Ground-based and Airborne Telescopes. Edited
  by Stepp, Larry M.. Proceedings of the SPIE, Volume 6267, pp. 626714

\bibitem[{{G{\"u}sten} {et~al.}(2006{\natexlab{b}}){G{\"u}sten}, {Nyman},
  {Schilke}, {Menten}, {Cesarsky}, \& {Booth}}]{guesten+2006b}
{G{\"u}sten}, R., {Nyman}, L.~{\AA}., {Schilke}, P., {et~al.}
  2006{\natexlab{b}}, \aap, 454, L13

\bibitem[{{Hatchell} {et~al.}(2005){Hatchell}, {Richer}, {Fuller},
  {Qualtrough}, {Ladd}, \& {Chandler}}]{hatchell2005}
{Hatchell}, J., {Richer}, J.~S., {Fuller}, G.~A., {et~al.} 2005, \aap, 440, 151

\bibitem[{{Helfand} {et~al.}(2006){Helfand}, {Becker}, {White}, {Fallon}, \&
  {Tuttle}}]{Helfand2006}
{Helfand}, D.~J., {Becker}, R.~H., {White}, R.~L., {Fallon}, A., \& {Tuttle},
  S. 2006, \aj, 131, 2525

\bibitem[{{Hildebrand}(1983)}]{hildebrand}
{Hildebrand}, R.~H. 1983, \qjras, 24, 267

\bibitem[{{Hinz} {et~al.}(2009){Hinz}, {Rieke}, {Yusef-Zadeh}, {Hewitt},
  {Balog}, \& {Block}}]{ref-hinz}
{Hinz}, J.~L., {Rieke}, G.~H., {Yusef-Zadeh}, F., {et~al.} 2009, \apjs, 181,
  227

\bibitem[{{Jackson} {et~al.}(2006){Jackson}, {Rathborne}, {Shah}, {Simon},
  {Bania}, {Clemens}, {Chambers}, {Johnson}, {Dormody}, {Lavoie}, \&
  {Heyer}}]{jackson+2006}
{Jackson}, J.~M., {Rathborne}, J.~M., {Shah}, R.~Y., {et~al.} 2006, \apjs, 163,
  145

\bibitem[{{Johnstone} {et~al.}(2004){Johnstone}, {Di Francesco}, \&
  {Kirk}}]{ref-johnstone}
{Johnstone}, D., {Di Francesco}, J., \& {Kirk}, H. 2004, \apjl, 611, L45

\bibitem[{{Lucas} {et~al.}(2008){Lucas}, {Hoare}, {Longmore}, {Schr{\"o}der},
  {Davis}, {Adamson}, {Bandyopadhyay}, {de Grijs}, {Smith}, {Gosling},
  {Mitchison}, {G{\'a}sp{\'a}r}, {Coe}, {Tamura}, {Parker}, {Irwin}, {Hambly},
  {Bryant}, {Collins}, {Cross}, {Evans}, {Gonzalez-Solares}, {Hodgkin},
  {Lewis}, {Read}, {Riello}, {Sutorius}, {Lawrence}, {Drew}, {Dye}, \&
  {Thompson}}]{Lucas2007}
{Lucas}, P.~W., {Hoare}, M.~G., {Longmore}, A., {et~al.} 2008, \mnras, 391, 136

\bibitem[{{Lumsden} {et~al.}(2002){Lumsden}, {Hoare}, {Oudmaijer}, \&
  {Richards}}]{lumsden+2002}
{Lumsden}, S.~L., {Hoare}, M.~G., {Oudmaijer}, R.~D., \& {Richards}, D. 2002,
  \mnras, 336, 621

\bibitem[{{Mac Low} \& {Klessen}(2004)}]{maclowklessen2004}
{Mac Low}, M.-M. \& {Klessen}, R.~S. 2004, Reviews of Modern Physics, 76, 125

\bibitem[{{Mauch} {et~al.}(2003){Mauch}, {Murphy}, {Buttery}, {Curran},
  {Hunstead}, {Piestrzynski}, {Robertson}, \& {Sadler}}]{ref-sumss}
{Mauch}, T., {Murphy}, T., {Buttery}, H.~J., {et~al.} 2003, \mnras, 342, 1117

\bibitem[{{McKee} \& {Ostriker}(2007)}]{McKee2007}
{McKee}, C.~F. \& {Ostriker}, E.~C. 2007, \araa, 45, 565

\bibitem[{{McKee} \& {Tan}(2002)}]{mckee-tan}
{McKee}, C.~F. \& {Tan}, J.~C. 2002, \nat, 416, 59

\bibitem[{{McKee} \& {Tan}(2003)}]{mckee-tan-2003}
{McKee}, C.~F. \& {Tan}, J.~C. 2003, \apj, 585, 850

\bibitem[{{Menten}(2007)}]{Menten2007}
{Menten}, K.~M. 2007, in IAU Symposium, Vol. 242, Astrophysical Masers and
  their Environments, Proceedings of the IAU, 496--505

\bibitem[{{Menten} {et~al.}(2005){Menten}, {Pillai}, \&
  {Wyrowski}}]{menten+2005}
{Menten}, K.~M., {Pillai}, T., \& {Wyrowski}, F. 2005, in IAU Symposium, Vol.
  227, Massive Star Birth: A Crossroads of Astrophysics, ed. R.~{Cesaroni},
  M.~{Felli}, E.~{Churchwell}, \& M.~{Walmsley}, 23--34

\bibitem[{{Molinari} {et~al.}(1996){Molinari}, {Brand}, {Cesaroni}, \&
  {Palla}}]{molinari+1996}
{Molinari}, S., {Brand}, J., {Cesaroni}, R., \& {Palla}, F. 1996, \aap, 308,
  573

\bibitem[{{Moore} {et~al.}(2007){Moore}, {Bretherton}, {Fujiyoshi}, {Ridge},
  {Allsopp}, {Hoare}, {Lumsden}, \& {Richer}}]{ref-moore}
{Moore}, T.~J.~T., {Bretherton}, D.~E., {Fujiyoshi}, T., {et~al.} 2007, \mnras,
  379, 663

\bibitem[{{Motte} {et~al.}(1998){Motte}, {Andre}, \& {Neri}}]{motte1998}
{Motte}, F., {Andre}, P., \& {Neri}, R. 1998, \aap, 336, 150

\bibitem[{{Motte} {et~al.}(2007){Motte}, {Bontemps}, {Schilke}, {Schneider},
  {Menten}, \& {Brogui{\`e}re}}]{motte+2007}
{Motte}, F., {Bontemps}, S., {Schilke}, P., {et~al.} 2007, \aap, 476, 1243

\bibitem[{{Motte} \& {Hennebelle}(2009)}]{motte2008}
{Motte}, F. \& {Hennebelle}, P. 2009, in EAS Publications Series, ed.
  L.~{Pagani} \& M.~{Gerin}, Vol.~34, 195--211

\bibitem[{{Motte} {et~al.}(2003){Motte}, {Schilke}, \& {Lis}}]{motte+2003}
{Motte}, F., {Schilke}, P., \& {Lis}, D.~C. 2003, \apj, 582, 277

\bibitem[{{Muders} {et~al.}(2006){Muders}, {Hafok}, {Wyrowski}, {Polehampton},
  {Belloche}, {K{\"o}nig}, {Schaaf}, {Schuller}, {Hatchell}, \& {van der
  Tak}}]{muders+2006}
{Muders}, D., {Hafok}, H., {Wyrowski}, F., {et~al.} 2006, \aap, 454, L25

\bibitem[{{Ossenkopf} \& {Henning}(1994)}]{ossenkopf+henning1994}
{Ossenkopf}, V. \& {Henning}, T. 1994, \aap, 291, 943

\bibitem[{{Palla} {et~al.}(1991){Palla}, {Brand}, {Comoretto}, {Felli}, \&
  {Cesaroni}}]{palla+1991}
{Palla}, F., {Brand}, J., {Comoretto}, G., {Felli}, M., \& {Cesaroni}, R. 1991,
  \aap, 246, 249

\bibitem[{{Palla} {et~al.}(1993){Palla}, {Cesaroni}, {Brand}, {Caselli},
  {Comoretto}, \& {Felli}}]{palla+1993}
{Palla}, F., {Cesaroni}, R., {Brand}, J., {et~al.} 1993, \aap, 280, 599

\bibitem[{{Pestalozzi} {et~al.}(2005){Pestalozzi}, {Minier}, \&
  {Booth}}]{pesta2005}
{Pestalozzi}, M.~R., {Minier}, V., \& {Booth}, R.~S. 2005, \aap, 432, 737

\bibitem[{{Price} {et~al.}(2001){Price}, {Egan}, {Carey}, {Mizuno}, \&
  {Kuchar}}]{Price2001}
{Price}, S.~D., {Egan}, M.~P., {Carey}, S.~J., {Mizuno}, D.~R., \& {Kuchar},
  T.~A. 2001, \aj, 121, 2819

\bibitem[{{Rathborne} {et~al.}(2006){Rathborne}, {Jackson}, \&
  {Simon}}]{rathborne+2006}
{Rathborne}, J.~M., {Jackson}, J.~M., \& {Simon}, R. 2006, \apj, 641, 389

\bibitem[{{Robitaille} {et~al.}(2008){Robitaille}, {Meade}, {Babler},
  {Whitney}, {Johnston}, {Indebetouw}, {Cohen}, {Povich}, {Sewilo}, {Benjamin},
  \& {Churchwell}}]{robitaille2008}
{Robitaille}, T.~P., {Meade}, M.~R., {Babler}, B.~L., {et~al.} 2008, \aj, 136,
  2413

\bibitem[{{Schilke} {et~al.}(1997){Schilke}, {Groesbeck}, {Blake}, \&
  {Phillips}}]{schilke+1997}
{Schilke}, P., {Groesbeck}, T.~D., {Blake}, G.~A., \& {Phillips}, T.~G. 1997,
  \apjs, 108, 301

\bibitem[{{Sewilo} {et~al.}(2004){Sewilo}, {Watson}, {Araya}, {Churchwell},
  {Hofner}, \& {Kurtz}}]{ref-sewilo}
{Sewilo}, M., {Watson}, C., {Araya}, E., {et~al.} 2004, \apjs, 154, 553

\bibitem[{{Shu} {et~al.}(1987){Shu}, {Adams}, \& {Lizano}}]{shu+1987}
{Shu}, F.~H., {Adams}, F.~C., \& {Lizano}, S. 1987, \araa, 25, 23

\bibitem[{{Simon} {et~al.}(2006){Simon}, {Jackson}, {Rathborne}, \&
  {Chambers}}]{simon+2006a}
{Simon}, R., {Jackson}, J.~M., {Rathborne}, J.~M., \& {Chambers}, E.~T. 2006,
  \apj, 639, 227

\bibitem[{{Siringo} {et~al.}(2009){Siringo}, {Kreysa}, {Kov{\'a}cs},
  {Schuller}, {Wei{\ss}}, {Esch}, {Gem{\"u}nd}, {Jethava}, {Lundershausen},
  {Colin}, {G{\"u}sten}, {Menten}, {Beelen}, {Bertoldi}, {Beeman}, \&
  {Haller}}]{siringo2008}
{Siringo}, G., {Kreysa}, E., {Kov{\'a}cs}, A., {et~al.} 2009, \aap, 497, 945

\bibitem[{{Siringo} {et~al.}(2007){Siringo}, {Weiss}, {Kreysa}, {Schuller},
  {Kovacs}, {Beelen}, {Esch}, {Gem{\"u}nd}, {Jethava}, {Lundershausen},
  {Menten}, {G{\"u}sten}, {Bertoldi}, {De Breuck}, {Nyman}, {Haller}, \&
  {Beeman}}]{siringo2007}
{Siringo}, G., {Weiss}, A., {Kreysa}, E., {et~al.} 2007, The Messenger, 129, 2

\bibitem[{{Sridharan} {et~al.}(2005){Sridharan}, {Beuther}, {Saito},
  {Wyrowski}, \& {Schilke}}]{sridharan+2005}
{Sridharan}, T.~K., {Beuther}, H., {Saito}, M., {Wyrowski}, F., \& {Schilke},
  P. 2005, \apjl, 634, L57

\bibitem[{{Sridharan} {et~al.}(2002){Sridharan}, {Beuther}, {Schilke},
  {Menten}, \& {Wyrowski}}]{sridharan+2002}
{Sridharan}, T.~K., {Beuther}, H., {Schilke}, P., {Menten}, K.~M., \&
  {Wyrowski}, F. 2002, \apj, 566, 931

\bibitem[{{Thompson} {et~al.}(2007){Thompson}, {Serjeant}, {Jenness}, {Scott},
  {Ashdown}, {Brunt}, {Butner}, {Chapin}, {Chrysostomou}, {Clark}, {Clements},
  {Collett}, {Coppin}, {Coulson}, {Dent}, {Economou}, {Evans}, {Friberg},
  {Fuller}, {Gibb}, {Greaves}, {Hatchell}, {Holland}, {Hudson}, {Ivison},
  {Jaffe}, {Joncas}, {Jones}, {Knapen}, {Leech}, {Mann}, {Matthews}, {Moore},
  {Mortier}, {Negrello}, {Nutter}, {Pestalozzi}, {Pope}, {Richer}, {Shipman},
  {Urquhart}, {Vaccari}, {Van Waerbeke}, {Viti}, {Weferling}, {White},
  {Wouterloot}, \& {Zhu}}]{ref-sassy}
{Thompson}, M.~A., {Serjeant}, S., {Jenness}, T., {et~al.} 2007, astro-ph,
  0704.3202

\bibitem[{{Urquhart} {et~al.}(2007){Urquhart}, {Busfield}, {Hoare}, {Lumsden},
  {Oudmaijer}, {Moore}, {Gibb}, {Purcell}, {Burton}, \&
  {Marechal}}]{Urquhart2007}
{Urquhart}, J.~S., {Busfield}, A.~L., {Hoare}, M.~G., {et~al.} 2007, \aap, 474,
  891

\bibitem[{{Vasyunina} {et~al.}(2009){Vasyunina}, {Linz}, {Henning}, {Stecklum},
  {Klose}, \& {Nyman}}]{ref-vasy}
{Vasyunina}, T., {Linz}, H., {Henning}, T., {et~al.} 2009, \aap, accepted,
  astro-ph/0902.1772

\bibitem[{{Walsh} {et~al.}(1997){Walsh}, {Hyland}, {Robinson}, \&
  {Burton}}]{walsh+1997}
{Walsh}, A.~J., {Hyland}, A.~R., {Robinson}, G., \& {Burton}, M.~G. 1997,
  \mnras, 291, 261

\bibitem[{{Ward-Thompson} {et~al.}(2007){Ward-Thompson}, {Andr{\'e}},
  {Crutcher}, {Johnstone}, {Onishi}, \& {Wilson}}]{ward-pp5}
{Ward-Thompson}, D., {Andr{\'e}}, P., {Crutcher}, R., {et~al.} 2007, in
  Protostars and Planets V, ed. B.~{Reipurth}, D.~{Jewitt}, \& K.~{Keil},
  33--46

\bibitem[{{Weingartner} \& {Draine}(2001)}]{weingartner}
{Weingartner}, J.~C. \& {Draine}, B.~T. 2001, \apj, 548, 296

\bibitem[{{Williams} {et~al.}(2000){Williams}, {Blitz}, \&
  {McKee}}]{williams+2000}
{Williams}, J.~P., {Blitz}, L., \& {McKee}, C.~F. 2000, Protostars and Planets
  IV, 97

\bibitem[{{Wood} \& {Churchwell}(1989)}]{wood+churchwell1989}
{Wood}, D.~O.~S. \& {Churchwell}, E. 1989, \apj, 340, 265

\bibitem[{{Wyrowski}(2008)}]{wyrowski2007}
{Wyrowski}, F. 2008, in Astronomical Society of the Pacific Conference Series,
  ed. H.~{Beuther}, H.~{Linz}, \& T.~{Henning}, Vol. 387, 3

\bibitem[{{Wyrowski} {et~al.}(2006){Wyrowski}, {Menten}, {Schilke},
  {Thorwirth}, {G{\"u}sten}, \& {Bergman}}]{wyrowski+2006}
{Wyrowski}, F., {Menten}, K.~M., {Schilke}, P., {et~al.} 2006, \aap, 454, L91

\bibitem[{{Young} {et~al.}(2006){Young}, {Enoch}, {Evans}, {Glenn}, {Sargent},
  {Huard}, {Aguirre}, {Golwala}, {Haig}, {Harvey}, {Laurent}, {Mauskopf}, \&
  {Sayers}}]{young+2006}
{Young}, K.~E., {Enoch}, M.~L., {Evans}, II, N.~J., {et~al.} 2006, \apj, 644,
  326

\bibitem[{{Zinnecker} \& {Yorke}(2007)}]{ZinneckerYorke}
{Zinnecker}, H. \& {Yorke}, H.~W. 2007, \araa, 45, 481

\end{thebibliography}

\end{document}